\documentclass{appolb}
\usepackage{epsfig}
\usepackage{cite}

\begin{document}
\eqsec

\title{Quasi-Particle Model  in Hot QCD
\thanks{ Based in part on lectures at the XLII Cracow School of Theoretical Physics, Zakopane, Poland May 30--June 8, 2003. The present version covers more material, and corrects some errors in the version published in the Proceedings of the School.}
}
\headtitle{Quasi-Particles  in Hot QCD}
\author{Chris P. Korthals Altes 
\address{Centre Physique Théorique au CNRS, Campus de Luminy\\ F13288, Marseille, Cedex,
France}}
\headauthor{C.P. Korthals Altes}
\maketitle

\begin{abstract}

Magnetic quasi-particles in hot QCD  predict Casimir scaling for the 
tension $\sigma$ of spatial Wilson loops. Recently this scaling has been verified on the lattice
to 1--2\% precision. We review how spatial loops tell us about the 
nature of the plasma state.  
 The strong first order nature for more than five colours and the fast drop of the topological susceptibilty at $T_c$
 found in simulations of hot quenched QCD are  consistent with a quasi-particle picture 
 valid down to the critical temperature. The magnetic quasi-particles will Bose condense in the ground state below $T_c$, and may  behave like a dilute Bose gas down to T=0. The model fixes the diluteness as $\sigma/m_M^2$. This is a small number for all temperatures.

\end{abstract}
\PACS{21.65+f,25.60+f}

\section{Introduction}

Plasmas are ubiquitous in the universe. Most of  visible matter is
in this state, in which atoms are fully ionized. A plasma we know quite
well is the sun. It sustains the thermonuclear power generation, that one tries
to imitate in Tokamaks. Large static magnetic fields show up on the sun's
surface as sunspots. They tell us that there is no magnetic screening
effect. Of course static long range electric fields are absent, due to Debye screening.

Today we are probing into a new plasma state, that of QCD, at RHIC. This plasma state features again Debye screening in the electric sector. But it is quite different in the magnetic sector. It has magnetic screening for all temperatures.
At low temperatures 't Hooft~\cite{thooftscreening} argued for a magnetic screening length. Such screening was expected on the basis of the dual superconductor analogy for QCD.  At high temperatures 
the need for a  magnetic screening length was
 recognized  also in the early days ~\cite{grosspis} of hot QCD.

It was based on the simple observation, that the static magnetic sector of QCD
is three-dimensional Yang--Mills theory and that this theory contains a mass gap.

This mass gap is, alas, not computable in terms of the small gauge coupling constant. But
it is accessible by lattice simulations. On the basis of these simulations we find that
the ratio of the square of the  magnetic screening mass and of the string tension of the square of the  magnetic screening mass is a
small number, about $1/20$ with small corrections of $O(1/N^2)$.

Some time ago a model for the three dimensional Yang--Mills theory was proposed~\cite{giovannakorthals},
in which precisely this ratio was supposed to be small. The model is based on the hypothesis that the partition
function can be approximately computed by thinking of the theory as
1being a dilute 3D gas of lumps with size equal to the magnetic screening length. The lumps
are non-perturbative quantities in terms of the 3D gluons. But they are supposed to be in
an adjoint magnetic ${\rm SU}(N)$ multiplet.

The merit of this model lies in  simple consequences for the tension of Wilson loops with
N-allity k. Their k dependence comes in through a simple multiplicity factor $k(N-k)$. Since then this dependence  has been verified to 1--2\%
by lattice simulations. Apart from the multiplicity factor the model predicts the tension
$\sigma$ of the loop to be simply the product of the magnetic screening $l_{\rm M}$ and the density $n_{\rm M}$ of the lumps:
\begin{equation}
\sigma\sim  l_{\rm M}n_{\rm M}\,.
\end{equation}

Multiplying this relation with $l_{\rm M}^2$ gives the desired diluteness $l_{\rm M}^3n_{\rm M}$.
This diluteness equals $l_{\rm M}^2\sigma$ and as remarked above, we know from 
 lattice simulations this
is a small number. Thus we have an a posteriori justification for the model.

In the next section~\ref{sec:debyeel} we recall some basic facts and numerology for the
Debye screening of electric charges, and how to simulate them on the lattice.
We introduce the electric flux loop in a  pedagogical manner.

In  section~\ref{sec:eff} we formulate the effective actions at high temperature. Those readers mostly interested in the quasi-particle picture can skip this section and the next~\ref{sec:critical} on the critical behaviour and universality. 
In the next section~\ref{sec:debyemag} we pursue the same but now for the magnetic
screening.

Finally we pass to the subject of magnetic quasi-particles
 in section~\ref{sec:model}.

The last section contains a brief discussion of the first order nature of the transtion at large number of colours. conclusions and prospects. Appendices do 
elaborate on some intricacies, especially on the use of a simple Stokes theorem.

\section{The Debye mass}\label{sec:debyeel}

Let us consider some gas heated at a temperature well above the ionization
energy. The electrons and ions are then moving more or less independently from each other.
 For simplicity we take the ions to have the opposite charge of the electrons.

  Since the ions
are much heavier than the electrons one can consider them to be a charged background with
a density $|e| n$. If you immerse a heavy point charge $Q$ in this
medium  its Coulomb interaction changes the density $n(\vec x)$ of the electrons around
it. The ions will remain unperturbed.  As all of you know this gives rise to screening of
the Coulomb law. The argument is purely classical. The Poisson equation in the presence of
the ionized electrons with charge $e$  reads:
\begin{equation}
\Delta A_0= e n(\vec x)- e n_i- Q \delta(\vec x)\,.
\label{poisson}
\end{equation}

The variation in the electron density is due to the variation in the
energy $eA_0$ of electrons in the field $A_0$, and hence in the corresponding Boltzmann factor
$\exp{(eA_0/T)}$( setting Boltzmann's constant equal to 1). Putting $A_0(\vec x=\infty)=0$ we
get $n(\vec x)=n\exp{(eA_0(\vec x)/T)}$. The Poisson equation becomes in the linear approximation in
the energy:

\begin{equation}
\Delta A_0= {e^2 n\over T}A_0- Q\delta(\vec x)\,.
\label{poissonlinear}
\end{equation}

Eq.~(\ref{poissonlinear}) is solved by a Yukawa potential,
$A_0=Q{\exp{-m_{\rm D}r}\over{4\pi  r}}$, with the Debye screening mass $m_{\rm D}$:
\begin{equation}
m_{\rm D}^2= {e^2 n\over T}\,.
\label{debyemass}
\end{equation}

The screening length $l_{\rm D}$ is the inverse of the screening mass $m_{\rm D}$.
Its raison d'\^etre is statistical, due to the Boltzmann factor. So
we expect the screening to involve many electrons. And this is precisely
what Eq.~(\ref{debyemass}) tells us: in a sphere of radius $l_{\rm D}$
we have  
\begin{equation}
l_D^3n=~T^{3/2}/(e^3n^{1/2})
\end{equation}
\noindent  electrons.

 To estimate this number we introduce
the atomic radius $r_a$ to split up the expression in two dimensionless factors:
\begin{equation}
l_D^3n=(T/(e^2/r_a))^{3/2}(r_a^3n)^{-1/2} 
\end{equation}

The first dimensionless factor is large because $T$ is larger than the
ionization energy $e^2/r_{a}$. The second dimensionless factor is $O(1)$ or larger because the number of electrons inside the atomic radius is certainly not larger in the ionized state.

 Typically, for the sun's corona $T/e^2$ is about $10^{2-3}$ the
atomic scale and the  number of electrons in the Debye sphere
is $10^6$. This condition is called the statistical screening or plasma condition.

It is useful to note that the thermal de Broglie wavelength $\lambda_B$ is small with respect to the interparticle distance. In fact the thermal wave length is on the order of the atomic radius and hence much smaller than the interparticle distance. The thermal wave length, just above ionization temperature, but well below pair creation temperatures is,
 in units of the atomic radius :
\begin{equation}
\lambda_B/r_a=h /(\sqrt{2\pi mT}r_a=\sqrt{2\pi} {\hbar\over{mvr_a}} {T\over{mc^2}}^{(-1/2)}{v\over c}\le O(1)
\end{equation}
We used that $v$, the electron velocity around the nucleus, is $\alpha$ times the velocity of light, and that the Bohr condition $\hbar /mvr_a$ is on the order of 1.

This estimate is valid for temperatures above ionisation but well below pair-creation scales. So  the  physics of the electron plasma just above ionization is classical. This is no longer true when pair creation becomes important.

\subsection{A Gedanken experiment}
It is amusing to do the following Gedanken experiment. It will provide us with
 a characterisation of the ionised state, or order parameter.
 Suppose we want to compute the
electric flux going through some large (with respect to the atomic size) closed loop $L$
with area $A(L)$. Normalize the flux ${\mit\Phi}=\int_{\rm S} d\vec S\cdot\vec E$ by the electron charge
$e$ and define :
\begin{equation}
V(L)=\exp{i2\pi{\mit\Phi}/e}\,.
\label{classicalfluxloop}
\end{equation}

  Of course, at $T$ below the ionization temperature no flux would be detected by the loop,
because there are only neutral atoms moving through the loop. 

Let us now raise the temperature above $T_{\rm ionisation}$. What will happen?
Both electrons and ions are screened. For simplicity we will take the ions to have the
opposite
of one electron charge.

We are going to make the following simplification. The charged particles are supposed to shine their flux through the loop if they are within distance $l_D$
from the minimal area of the loop. This defines a slab of thickness $2l_D$. 

Of course if we plot $|{\mit\Phi}/e|$ as function of
the distance of the particle to the loop you find an exponential curve with the maximum
$1/2$ at zero distance. For the sake of the argument we will replace that curve by a theta
function of height $1/2$ and width $2l_{\rm D}$. If one wants to do better one has to
deal with infinitesimally thin slabs, and integrate over the thickness, and we will do so in Appendix A.
The result is parametrically the same as the one we will derive keeping the simple minded method.

Then one electron (ion) on the down side of the loop 
will
contribute $+1/2 (-1/2)$ to the flux, and with opposite sign if on the up side of the
loop. That is: $V(L)|_{\rm one\;charge}=-1$. This result is independent of the sign of the charge! The plasma is overall neutral, the loop is sensitive only to charge fluctuations. For $l$ charges inside the slab the flux adds linearly and  we find:
\begin{equation}
V(L)|_l=(-1)^l.
\label{lcharges}
\end{equation}

Assuming that all charges move independently,  the average of the flux
loop $V(L)$ is determined by the probability $P(l)$ that $l$ electrons (ions)
are present in the slab of thickness $2l_{\rm D}$ around the area spanned by the loop.  Taking
for $P(l)$ the Poisson distribution ${1\over {l!}}(\bar l)^l\exp{-\bar l}$-- $\bar l$ is
the average number of electrons (ions) in the slab--  we find for the thermal average of the
loop:
\begin{equation}
\langle V(L)\rangle_T=\sum_lP(l)V(L)|_l=\sum_l P(l)(-)^l=\exp{-4\bar
l}\,.
\label{electrictension}
\end{equation}

Now $\bar l=A(L)2l_{\rm D}n(T)$, so the electric flux loop obeys an area law
 $\exp{-\rho(T)A(L)}$, with a tension 
\begin{equation}
\rho(T)=8l_{\rm D}n(T).
\label{classicaltension}
\end{equation}
 Note that the 
tension in units of the Debye mass, $l_{\rm D}^2\rho$, equals the number of electrons in a Debye volume $ 8l_{\rm D}^3n(T)$, because of eq.(\ref{classicaltension}). We saw already that this number was huge.

A caveat  behooves the word ``tension''.  On the minimal area the density of
electrons is not changing, only near the perimeter. Nor is the free energy density. So the word tension is not the correct one, though we will continue to use it.

Another  aspect of this tension is that no quantum theory came into
the calculation. The only ingredient that takes into account the dynamics of the plasma is the screening length $l_{\rm D}$ and the statistical independence
through use of the Poisson distribution.

In conclusion the behaviour of the loop is very different for the ionised state, as compared to the de-ionised state.  It behaves with an
area law above ionization temperature. Un-ionized atoms can at most cause perimeter effects.

\subsection{A closer look at the flux loop} 

 Untill now only classical physics went into
the calculation of the flux loop average. 

If we assume for the electron density $n$ the Fermi-Dirac distribution
our calculation is semi-classical. But only at a  temperature where relativistic effects are starting to operate ($T\sim 2m_e$) the difference between the Poisson distribution and Fermi-Dirac becomes important.  For temperatures, where electrons and positrons become easily pair produced we need  a  fully-fledged field theory. As we will see at the ned of this section we can compute $\rho$  in a WKB approximation. Such approximations are of $O(1/g)$ and are therefore large. 

 And our Gedanken experiment takes
in field theory a perhaps surprising form. The flux $\int_{\rm L} d\vec S\cdot\vec E$  becomes now an operator, $\vec E$ being  the canonical conjugate of the vector potential $\vec A$, with $[E_k,A_l]={1\over i}\hbar  \delta_{k,l}$, the Kronecker delta standing for a delta function of the space arguments as well. 

The loop acts on the state space, and  in particular we are interested in what it does to physical states. What are physical states? First we should observe that at the ambient temperature $T\ge 2m_e$ the nuclei start to come apart. So we have a field theory of electrons, positrons, and nucleons.
A physical state is one where a given state contaning a set of particles is
made gauge invariant. How to do this can be seen from  a gauge transformation on the photon field:
\begin{equation}
A_k\rightarrow A_k^{\omega}= A_k+{1\over e}\partial_k\omega=A_k-{1\over {ie}}\exp{i\omega}\partial_k\exp{-i\omega}
\label{photongauge}
\end{equation}
and for the matter fields:
\begin{equation}
\psi\rightarrow \exp{(i\omega)}\psi .
\label{matgauge}
\end{equation}

So the gauge group is $U(1)$, and $\omega$ is the angle characterizing the U(1)
phase $\exp{i\omega}$ lying on the unit circle. 
Now a physical state is produced by averaging a given state 
over all possible
gauge configurations $\omega$. ``All'' excludes those  configurations where you can find a closed curve along which
$\omega$ winds around the unit circle once or more. That is to say, all admitted
gauge transformations  constraining the physical Hilbert space are in the trivial homotopy class $\Pi_1$ of the unit circle $U(1)$~\footnote{Had we constrained the physical Hilbertspace even more, by including transformations in the non-trivial homotopy classes, the loop $V(L)$ would have had no physical effect.}.

To understand the action of the flux operator $V(L)$ we consider a gauge transformation $\omega_1(L)$. The suffix indicates a discontinuity of $2\pi$ when we go with the  gauge transformation through the area spanned by the loop $L$.
One can take for the gauge function  half the solid angle spanned by the loop. This makes a jump of $2\pi$ at the surface, and vanishes at infinity.

Gauss' theorem tells us then: 
\begin{equation}
\int d\vec S.\vec E ={1\over{2\pi}}\int dV\vec\nabla(\vec E\omega_1(L)) 
\label{mattergauge}
\end{equation}

\noindent with the volume integral over all space.
This formula is true for any $\omega_1(L)$ with discontinuity $2\pi$ on the surface $S(L)$, and which falls off sufficiently fast at spatial infinity, so that there we are allowed to drop the surface integral. 
The r.h.s. can be rewritten with the help of the charge density operator $ej_0$ as:
\begin{equation}
{1\over{2\pi}}\int dV\vec\nabla(\vec E\omega_1(L))={1\over{2\pi}}\int dV[(\vec\nabla.\vec E-ej_0)\omega_1(L)+(\vec E.\vec\nabla+ej_0)\omega_1(L)]
\end{equation}
So the flux operator can be written on the physical subspace ( where \linebreak  $\nabla.\vec E-ej_0=0$) as  a gauge transformation with the singular gauge function.

\begin{equation}
V(L)=\exp{\{i{1\over e}\int dV(\vec E.\vec\nabla+ej_0)\omega_1\}}.
\label{discgauge}
\end{equation}

According to this formula  $V(L)$ creates a magnetic flux line along the closed loop
$L$. Everywhere else it is just a regular gauge transformation.
 So its physical effect  is localized on the loop $L$. Hence the volume integral
in eq. (\ref{discgauge}) reduces to a line integral along $L$, that we can write formally as:
\begin{equation}
V(L)=\exp{i{1\over e}\int dV(\vec E.\vec\nabla+ej_0)\omega_1}=\exp{i\int_L d\vec l.\vec \mu}.
\label{magneticline}
\end{equation}

 The flux $d\vec l.\vec \mu$  has strength
 $\oint_{\gamma} d\vec s.{1\over e}\vec\nabla\omega_1={2\pi\over e}$, where $\gamma$ is any closed curve encircling $L$ once. This means that a Wilsonloop:
\begin{equation}
W(\gamma)=\exp{i{e\over{\hbar}}\oint_{\gamma}d\vec s.\vec A}
\end{equation}
will commute with the electric flux loop. And this is borne out by using the canonical commutation relations.

 It is amusing that in the process
we have uncovered a dual  Stokes theorem that relates $V(L)$ as line integral of this magnetic flux to $V(L)$ as  surface integral of the electric field strength. It is only valid in the physical subspace.



Loops with an integer multiple $k$ of unit flux are possible too. They will be
  denoted by $V_k(L)$ and correspond to the n-th homotopy class of the homotopy group $\Pi_1(U(1))$. Our ''dual Stokes theorem'' then tells us that   

\begin{equation}
V_k(L)=\exp{i{2\pi\over e} k\int d\vec S.\vec E}.
\end{equation}

This concludes our simple-minded  exploration of the flux properties of the plasma. The question is: can we do better?
To answer that question we have to use a more sophisticated version of our screened electron gas.

\subsection{The Coulomb gas}\label{subsec:coulomb}

One reproach one can make to the model in the previous section is its treatment
of the thermal fluctuations. The only way they are built in is through
  the Debye mass and the Poisson distribution. To take them more fully into account we are going to analyse the three dimensional Coulomb gas. This model still ignores the time dependent fluctuations in QED. 

 The Coulomb gas has as degrees of freedom the density $n(\vec x)=\sum_{\vec x_i}n_i\delta(\vec x - \vec x_i)$. $n_i$ is the sum of all the charges in the point $\vec x_i$. The interaction is by means of the Coulomb propagator:
\begin{equation}
D(\vec x-\vec y)=\langle\vec x|{1\over {-\nabla}^2}|\vec y\rangle
\end{equation} 
The partition function is then given by:  
\begin{equation}
Z=\sum_{n(\vec x)}\exp{-{e^2\over {2T}}\int_{\vec x,\vec y}n(\vec x)D(\vec x-\vec y)n(\vec y)- \mu^3 \int_{\vec x}n(\vec x)^2}.
\label{coulombpartition}
\end{equation} 

The parameter $\mu$ has dimension of mass, and serves to control the charge density.

We introduce a continuous parameter $A(\vec x)$ that decouples the densities in
the partition function. In what follows we are going to simplify notation by replacing integrals by dots and dropping the position vectors. So we have:

\begin{equation}
Z=\int DA\sum_{n}\exp{\{-{1\over 2}(({T\over {e^2}})\nabla A.\nabla A)-in.A-
\mu^3  n^2\}}.
\end{equation}

 Whatever the value of $\mu$, after summing over $n$ we find an effective 
potential $V(A)$,
 periodic  mod $2\pi$ in $A$:
\begin{equation}
Z=\int DA\exp{-\int {1\over 2}({T\over {e^2}}(\nabla A)^2+V(A)}.
\label{periodicpot}
\end{equation}

Apart from periodic, the potential is even in $A$. So it can be written as a sum of cosines. For small $\mu$ only the $\cos A$ term appears. So for small $A$
we get a mass term $e^2\mu^3/T$, so the $A$ field is the screened scalar potential.

Let us finally compute the average of the electric flux loop in this theory.
Remember the flux loop can be formulated as any gauge transform in three dimensional  space, where the gauge parameter $\omega$ makes a jump of $2\pi$ at the surface. The gauge parameter  is nothing but
our variable $A(\vec x)$!

\begin{figure}
\includegraphics[angle=270]{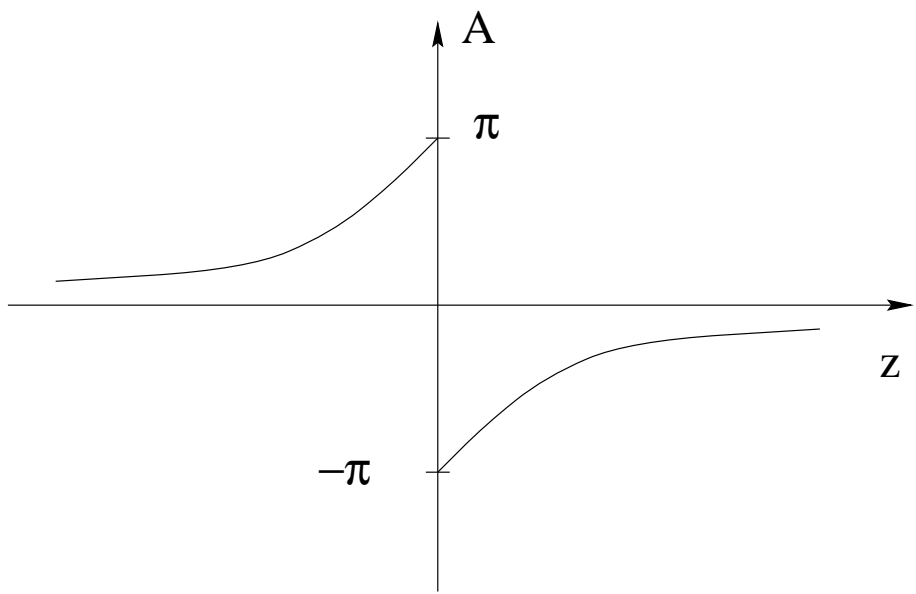}
\label{fig:profile}
\caption{The profile of the field $A$ that minimizes
 the effective action in eq.(\ref{periodicpot}).}
\end{figure}

 Thus, the average of the loop is determined from the
effective action in eq.(\ref{periodicpot}), where $A$ now depends only on the coordinate normal to the loop surface, vanishes at infinity
and makes a jump of $2\pi$ at the surface--see fig.(1)--:

\begin{equation}
<V_1(L)>=\int DA\exp{-L_{tr}^2\int dz ({T\over {2e^2}}(\partial_z A)^2+V(A))}.
\end{equation}
As the transverse size of the loop, $L_{tr}\rightarrow\infty$ we only need the minimum of the effective action. 

Since the potential $V$ is periodic mod $2\pi$, we can  lift the branch on the right hand side of the loop area in fig.(1) over $2\pi$. Then we have a continuous profile starting at $A=0$ at $-\infty$ and ending at $A=2\pi$ at $+\infty$.

This is the solution of the classical problem where the profile $A(z)$ rolls from one maximum to the next one in $V$. Again we found an area law for the loop,
the tension being
\begin{equation}
\rho_1=min_{A}\{\int dz ({T\over {2e^2}}(\partial_z A)^2+V(A))\}.
\end{equation}

Note that the potential V(A) does not depend on the coupling $e$.
One has to absorb the coupling $e$ into the z-coordinate, in order to 
balance kinetic and potential energy: the profile that minimizes the 
action must be of the form $A(ez)$. A scaling argument then leads to
\begin{equation}
\rho_1=c{1\over e}T^2.
\end{equation}

The precise knowledge of $V$ does not matter for the parametric form. The numerical constant $c$ reflects the ratio of $T$ and $\mu$, and the form of the potential.
The reader recognizes the same parametric form as in the result eq.(\ref{classicaltension}) from the quasi-particle picture.

\subsection{Summing up}
Let us conclude his long section with a short summary. The quasi-particle picture gives an area law for the loop, just from the Poisson distribution. A more sophisticated version, the Coulomb gas, gives an area law as well. The new feature is its  semi-classical nature: it is due to tunneling through an effective potential barrier.

If we want to extend the calculation to field theory, with time dependent fluctuations, we have to worry about $A_0$, the variable in our effective potential,
since it is no longer gauge invariant. This is naturally mended by replacing
the static $A_0$  by the phase of the Polyakov loop or Wilson line:
\begin{equation}
P(A_0)={\cal{P}}\exp{ie\int^{1/T}_0 A_0(\tau,\vec x))d\tau}\equiv \exp{iA}.
\end{equation}

So in the full gauge theory we are supposed to compute the effective potential as a function of this phase. This can be done in a saddle point approximation around the phase $A$.    Once done, the calculation  becomes the minimization of that action. 

To one loop order the potential $V$ equals the logarithm of the determinant of the fluctuations around $A$.  This determinant has been studied extensively in the litterature~\cite{bhatta} and the result for the tension is parametrically the same as what quasi-particles gave us. In terms of $q\equiv {A\over{2\pi}}$
and the rescaled z-variable $z^{\prime}=eTz$ one finds for the effective action :
\begin{equation}
S_{eff}={2\pi^2T^2\over e}\int dz'\big((\partial_{z^{\prime}}q)^2+{2\over 3}({1\over{16}}-(q+{1\over 2})^2(1-|q+{1\over 2}|)^2)\big)
\end{equation}

The potential $V(q)={2\over 3}({1\over{16}}-(q+{1\over 2})^2(1-|q+{1\over 2}|)^2)$ is the logarithm of the fermionic determinant in the background $q$ normalized by the determinant without $q$.

The minimum of this action is easily found by completing the square of the kinetic term: 
$$(\partial_{z^{\prime}}q)^2+V(q)=(\partial_{z^{\prime}}q-\sqrt{V(q)})^2+2\partial_{z^{\prime}}q\sqrt{V(q)}.$$

The minimum is realized  when the square cancels. This gives us the equation of motion
for  the profile $q$. The remaining term is the actual value for $\rho_1$ and equals 
$2\pi^2T^2\int dq\sqrt{V(q)}$ through change of variables.

Rewrite this as $c_fl_Dn$, using the lowest order result $l_D^{-2}= {e^2T^2\over 3}$ and $n=0.87...( {T^3\over{2\pi^2}})$ , and you have the same parametric result as that from the quasi-particle picture.  

 Only the numerical factor $c_f$ is quite a bit larger than in 
eq.(\ref{classicaltension}), because the time-dependent fluctuations 
have now been taken into account.

\section{Effective field theories at high temperature}\label{sec:eff}

With what we have learnt from the static approximation in mind we turn to QCD at high $T$.

Any field theory in equilibrium at non-zero temperature can be formulated
as a Euclidean path integral. The time direction in that integral is periodic mod $1/T$
for bosons, and anti-periodic for fermions. For the statistical Gibbs sum one has
\begin{equation}
\Tr_{\rm phys}\exp{-H/T}=\int DA DqD\bar q\exp{-{1\over{g^2}}S(A,q,\bar q)}\,,
\label{gibbsfeynman}
\end{equation}
\noindent where the gauge potentials $A$ and the quark fields $q$ are integrated over.
Under gauge transformations $\Omega$ we have:
\begin{eqnarray}
A&\rightarrow& \Omega A\Omega^{\dagger}-{1\over{ig}}\Omega\partial\Omega\\
q&\rightarrow& \Omega q
\label{convention}
\end{eqnarray}
\noindent and covariant derivatives $D(A)=\partial-ig[A,$ and $\nabla=\partial-igA$. From the commutator of the partial derivatives follows $F_{\mu,\nu}=\partial_{\mu}A_{\nu}-\partial_{\nu}A_{\mu}-ig[A_{\mu},A_{\nu}]$.

In the limit that $T$ becomes small with respect to typical particle scales
the time direction can be neglected. This is called dimensional reduction~\cite{pis}.
It can be formulated as a systematic approximation scheme using
 that  QCD at high temperature $T$ has a small running coupling $g(T)$.
The inverse propagator of a boson is proportional to $(2\pi nT)^2+{\vec p}^2$.
Here $n$ takes on integer values.  For a fermion $n$ is replaced by $n+1/2$.
As $T$ becomes large one can integrate all heavy $T$-modes (``hard'' modes) and stay with
a three dimensional theory in terms of only the static bosonic modes with $n=0$.
For QCD this 3d Lagrangian reads:
 \begin{eqnarray}
{\cal{L}}_{\rm E} & = & {\rm Tr}(\vec D(A)A_0)^2+m_{\rm E}^2{\rm Tr}A_0^2+\lambda_{\rm E}({\rm Tr}(A_0^2))^2 \nonumber\\
              &&+  \bar\lambda_{\rm E} \big({\rm Tr}(A_0)^4-{1\over 2}({\rm Tr}A_0^2)^2\big)+{\rm Tr}
F_{ij}^2+\delta{\cal{L}}_{\rm E}\,.
\label{estat}
\end{eqnarray}

This is called the electrostatic Lagrangian.

The last term contains higher powers of $A_0$ and of the covariant derivative
$\vec D(A)=\vec\partial-ig_{\rm E}[\vec A$. Neglecting it means one neglects $O(g^4)$ in the
correlations you compute with the first six terms. The  parameters in this Lagrangian  are
computed from the corresponding n-point functions in one and two loop accuracy in terms of
$g_{\rm E}^2=g^2(T)T$. Higher loop order adds only to the accuracy if one takes into account
$\delta{\cal{L}}_{\rm E}$.
The electrostatic coupling $g^2_{\rm E}$ is to one loop order in terms of
${\mit\Lambda}_{\overline{MS}}$ (no flavours):
\begin{equation}
{g_{\rm E}
^2N\over{T}}={24\pi^2\over{11\log({6.742.T\over{{\mit\Lambda}_{\overline{MS}}}})}}\,.
\label{minsub}
\end{equation}

The subtraction was chosen to minimize the one loop effects~\cite{huang}.

We have swept one problem under the rug. When integrating the hard modes we have to admit
a lower cut-off ${\mit\Lambda}_{\rm E}$, in between the scale $T$ and the electrostatic scale $gT$. In
principle the parameters will depend on this cut-off.

One expects that we have the same picture as before: above the deconfining
temperature $T_{\rm c}\sim 200 ~{\rm MeV}$) we have a gas of ``ions'', the quarks, and of
``electrons'', the gluons. There is screening as before, as witnessed by the mass
parameter $m_{\rm E}^2={N\over 3}g^2T^2$ in electrostatic Lagrangian, Eq.~(\ref{estat}).

This constitutes the Stephan-Boltzmann picture of QCD and  interactions between gluons and
quarks describe deviations from
this free quasi-particle system.

Specific to QCD is that there is not only an electrostatic scale set by the Debye mass.
We can integrate out in electrostatic Lagrangian all degrees of freedom corresponding to
the Debye scale. That leaves us with the magnetostatic Lagrangian:
\begin{equation}
{\cal{L}}_{\rm M}={\rm Tr}F_{ij}^2+\delta{\cal{L}}_{\rm M}
\label{eq:mstat}
\end{equation}
\noindent
with a magnetostatic gauge coupling $g^2_{\rm M}$ in $F_{ij}$.

Here an ultra-violet cut-off ${\mit\Lambda}_{\rm M}$ is needed. It separates the electrostatic scale
$gT$ and the magnetic scale $g^2T$.

Terms with higher order covariant derivatives are contained in $ \delta{\cal{L}}_{\rm M}$. They
are needed when we want an accuracy of $O(g^3)$.

The magnetic action gives a non-perturbative theory. In calculating a Green's
function with a typical external momentum $p$ we find as dimensionless parameter $g_{\rm M}^2/p$
which is $O(1)$ if the momentum is  the magnetic scale. In particular calculation of the
free energy in this theory will give for dimensional reasons $(g^2_{\rm M})^3$ times a
non-perturbative constant. That is the contribution one
expects from a 4-loop diagram. As for the Green's function all higher loops
are of the same order.

Still we can compute a series in the small coupling $g(T)$. Only the coefficients are
non-perturbative.

A useful comment on the parameters in the two effective actions is in order ~\footnote{Brought up in conversations with M. Laine.}.
Their determination is perturbative. Truncation of higher derivative terms 
in the electro- (magneto-)static actions meant dropping terms of relative order $O(g^4)$ ($O(g^3)$).
This means we need to calculate the parameters to these precisions. In particular we need $g_E$ to one loop order as in eq.(\ref{minsub}), and $g_M$ to two loop order in terms of $g_E$ and $m_E$. The latter relation between magnetic and electric scales by integrating the out the electric scales from the electrostatic action and matching with the magnetostatic action  has been computed and gives~\cite{giovannamag}
\begin{equation}
g^2_MN=g^2_EN\big(1-{1\over {48}}{g^2_EN\over {\pi m_E}}-{17\over{4608}}({g^2_EN\over {\pi m_E}})^2+O(g^3)\big).
\label{elmag2}
\end{equation}
As $m_E$ is of order $g$ the dimensionless expansion parameter is $O(g)$.

For asymptotically large temperatures such a picture is indeed accurate.
But asymptotic means temperatures about $10^6 T_{\rm c}$, well above the electro-weak scale, far away from where RHIC physics operates .
It turns out, however, that the accuracy varies with the observable.

In this subsection we will take the pressure as an example, how the expansion works.
 To put the calculation of the contributions of order higher than three in perspective and
to see how the different scales come in, we recall once more
the hierarchy of scales, cut-offs ${\mit\Lambda}$ and reduced actions needed to compute the
pressure:
$$T\gg{\mit\Lambda}_{\rm E}\gg gT\gg
   {\mit\Lambda}_{\rm M}\gg g^2T\,.$$

The pressure is normalized by $p_0=p_{\rm Stefan-Boltzmann}$ and consists of three parts:
$${p\over{p_0}}=p_{\rm h}+p_{\rm E}+p_{\rm M}\,.$$
The hard modes are cut-off in the infrared by ${\mit\Lambda}_{\rm E}$ and equal $p_{\rm h}$.
Schematically we get:
$$p_{\rm h}=1+g^2+g^4\log{T\over{ {\mit\Lambda}_{\rm E}}}+g^4+g^6log{T\over{{\mit\Lambda}_{\rm E}}}+g^6+...\;.$$

All powers of the coupling are even, since infrared divergencies are cut-off by
${\mit\Lambda}_{\rm E}$. The short distance scales (larger than $T$)
are absorbed in the running coupling, Eq.~(\ref{minsub}). The cut-off ${\mit\Lambda}_{\rm E}$ appears
only in logarithms.
The electric mode contributions are computed with ${\cal{L}}_{\rm E}$ and give $p_{\rm E}$:
$$p_{\rm E}=g^3+g^4\log{{\mit\Lambda}_{\rm E}\over{m_{\rm E}}}
+g^4+g^5+g^6\log{{\mit\Lambda}_{\rm E}\over{m_{\rm E}}}+g^6\log{m_{\rm E}\over{{\mit\Lambda}_{\rm M}}}+g^6+...\;.$$

Note the odd powers in $g$. They come in because the electric mass $gT$ comes
in through propagators from the electrostatic action, Eq.~(\ref{estat}).
For example, the contribution from the scalar potential $A_0$ gives the first term in
$p_{\rm E}$:
\begin{equation}
-{1\over 2}(N^2-1)\int {d\vec l\over{(2\pi)^3}}\log(\vec k^2+m_{\rm E}^2)={{\mit\Gamma}(-{3\over
2})\over{16\pi^{{3\over 2}}}}m_{\rm E}^3\,.
\label{eq:cubic}
\end{equation}

The dominant cubic term was computed in Eq.~(\ref{eq:cubic}). We can expect
 logarithms of the two ratios of the three scales $m_{\rm E}$, ${\mit\Lambda}_{\rm E}$
  in the electrostatic action and  ${\mit\Lambda}_{\rm M}$.

Finally the  magnetic contribution is computed with ${\cal{L}}_{\rm M}$:
 $$p_{\rm M}=g^6\log{{\mit\Lambda}_{\rm M}\over {g^2_{\rm M}}}+g^6+...\;.$$

We only put in the obvious dependence on the parameters in the electrostatic and
magnetostatic actions. There are three comments:
\begin{itemize}
\item
All terms shown are perturbatively calculable, except the last one in $p_{\rm M}$.
\item
All perturbatively calculable terms have been computed~\cite{zhai}, except for the $g^6$
terms. In particular the log's are known by now~\cite{log}.
\item
All dependence on the cut-offs cancels, as expected.
\end{itemize}

\begin{figure}[htb]
\centerline{%
    \epsfig{file=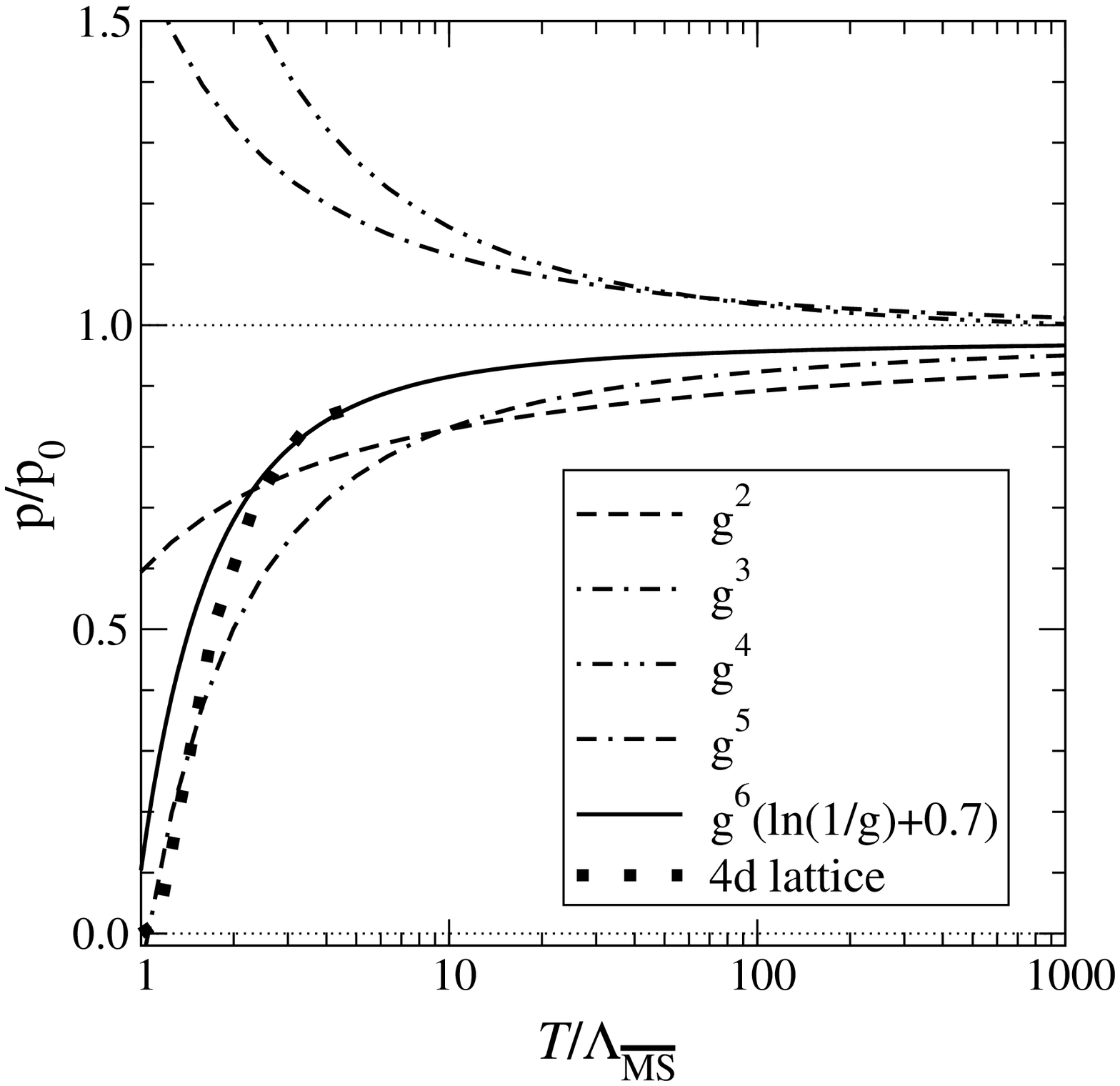,width=6cm} 
    \epsfig{file=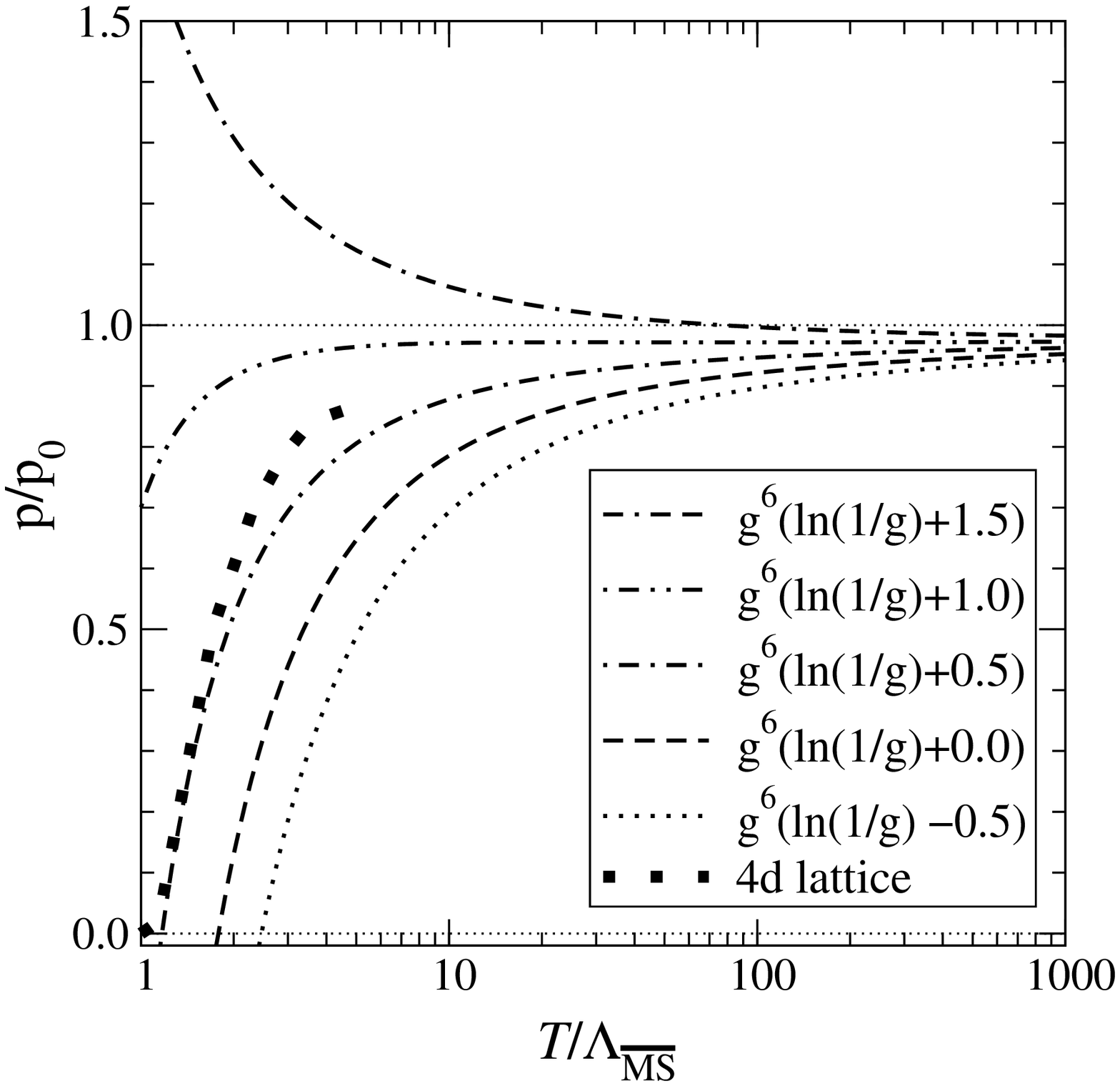,width=6cm}}
\caption[a]{Left: perturbative results at various orders for pure SU(3) gauge theory,
including
${\cal O}(g^6)$ for an optimal constant. Right: the dependence of the
${\cal O}(g^6)$ result on the (not yet computed) constant, which
contains both perturbative and non-perturbative contributions.
The 4d lattice results are from~\cite{boyd}. From Ref.~\cite{log}.}
\label{fig:pert_g6}
\end{figure}
This  is dramatically illustrated by Fig. \ref{fig:pert_g6}. You see on the left
the lattice data for the pressure in units of its Stephan--Boltzmann value
${8\pi^2\over{45}}T^4$ plotted together with the known low order (up to $O(g^6)$)
perturbative results.
For the asymptotically large temperatures mentioned the quasi-particle  picture is indeed
accurate as the figure shows.
The right panel shows how the prediction can improve, when the known~\cite{log}
logarithmic contribution to the $O(g^6)$ coefficient is included, together with a guess
for the non-perturbative part of the coefficient.

For any reasonable
$T$, say below 2 GeV, the coupling obeys $g^2_{\rm E}(T)/$\break $T\le 2.5$. This is about 30 times
bigger then $e^2$, so we may already surmise that low order perturbation theory will be
far from accurate.

 \subsection{The Debye mass and electric flux loop in QED and QCD}\label{subsec:debyemass}

Now we discuss the Debye screening in the QCD plasma.

\begin {figure}[htb]
\centerline{%
      \epsfig{file=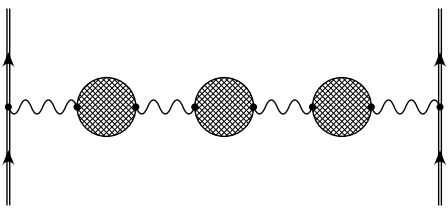,width=6cm}}
   \caption
       {%
       A single gluon exchanged between two static test charges.}
       \label{figa}
\end {figure}

Let us put a probe charge in the plasma, say a very heavy quark.
In Fig.~\ref{figa} the exchange of a single gluon  is shown, together
with its multiply\break

\begin{figure}[htb]
\centerline{%
      \epsfig{file=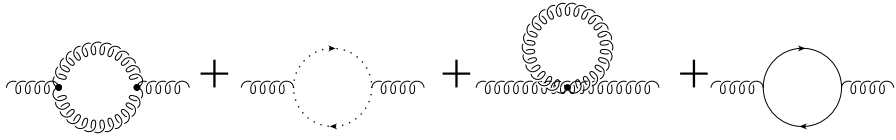,width=12cm}}
   \caption
       {%
       One-loop self-energy of a gluon.
       \label{figc}
       }%
\end {figure}

\noindent inserted self-energy. Once    we compute the self energy
${\mit\Pi}_{00}$ of the gluon, shown in Fig.~\ref{figc}, the resumed propagator $D_{00}$
becomes:
\begin{equation}
D_{00}(\vec p)={1\over{{\vec p}^2+{\mit\Pi}_{00}(\vec p)}}\,.
\end{equation}

It is easy to see that only the hard modes contribute to the self energy.

 Hence one finds the one loop result for $m_{\rm E}^2$ in the electrostatic Lagrangian.
\begin{equation}
D_{00}(r)\sim {\exp{-m_{\rm E} r}\over r}\,.
\end{equation}
 To two loop order one finds that  already at that order  non-perturbative effects
contribute.

Hence a definition independent of perturbation theory is called for, and a natural
candidate
 is the correlator between two heavy test
charges: its fall-off as a function of distance gives us the screening mass.

The test charge put in the plasma changes the free energy. This change can be
expressed in terms of an expectation value of the thermal Wilson line:
\begin{equation}
\exp{-{{\Delta F_{\psi}}\over T}}={\int DA{1\over N}\mbox{Tr}{\cal{P}}(A_0)\exp{-S(A)}\over {\int DA
\exp{-S(A)}}}\equiv\langle P(A_0)\rangle
\label{excess}
\end{equation}
where the thermal Wilson line is given by:
\begin{equation}
P(A_0(\vec x))={1\over N}\mbox{Tr}{\cal{P}}\exp{ig\int\limits_0^{1\over T}d\tau A_0(\vec
x,\tau)}\,.
\label{wilsonline}
\end {equation}

If the test charge is in the fundamental representation then so is $A_0$.

The path ordering is defined by dividing the interval $[0,1/T]$
into a large number  $N_{\tau}$ of bits of length $\Delta\tau={1\over {N_{\tau}T}}$:
\begin{equation}
{\cal{P}}(A_0)=\lim_{N_{\tau}\rightarrow\infty} U(\tau=0,\Delta\tau)U(\tau=\Delta\tau
,2\Delta\tau)\ldots U\left(\tau={1\over T}-\Delta\tau,{1\over T}\right)\,.
\label{wilsonlineordering}
\end {equation}

>From a formal and from a computational point of view the correlator $\langle
P(A_0)P(A_0)^{\dagger}\rangle$ has two advantages over the calculation presented above
using the scalar potential:

\begin{itemize}
\item
The correlator is gauge invariant.
\item
The correlator can be evaluated non-perturbatively, \ie on the lattice.
\end{itemize}

Both are needed for an accurate determination of the screening length
in QCD. Perturbation theory is not enough, despite the small coupling
$g(T)$ for asymptotically large temperature (\ie well above the electro-weak scale).
In the confined phase
the correlator $\langle P(A_0(r))P(A_0(0))^{\dagger}\rangle$ obeys an area law $\exp{-\sigma {r\over
T}}$, in the deconfined phase the area law is replaced by the Yukawa potential controlled
by the Debye mass.

There is a further advantage: correlators of gauge invariant operators will excite the
levels of a fictitious Hamiltonian describing Yang--Mills dynamics in a space with one
periodic mod $1/T$ direction and two other infinite directions. The time conjugate to this
Hamiltonian is now the direction of the correlation.

Conserved quantum numbers are then, apart from those from the two dimensional rotation
group, the usual discrete parities, charge conjugation $C$, parity $P$ (now in 2D),
and a new quantum number, that changes $A_0\rightarrow -A_0$, called $R$-parity.
The magnetic quasi-particles will condense in the ground state below $T_c$. 

Note that the Debye mass defined this way should coincide in one loop order with what we
found before: $m_{\rm E}$ in the electrostatic Lagrangian.
So it is associated with $R$-parity $-1$. Clearly, this corresponds to
the imaginary part of the Wilson line. The real part excites $R=+,\;P=C=+$ states. To wit:
if the correlator $\langle PP\rangle$ between two like charges were zero, then
the difference between correlators of imaginary and real parts would be zero.
That would mean, in turn, that the masses controlling their decays would be degenerate.
This is not the case. Two like screened charges are compatible on a
torus. In the confined phase (no screening) their correlation is indeed zero.

\section{Z(N) symmetry, universality, and the order of the transition}\label{sec:critical}

Let us move from the very high $T$ region to the critical region near $T_c$.
There is a symmetry due to invariance of the Yang--Mills action under
gauge transformations that are not periodic  in Euclidean time, but only periodic modulo a
center group element $\exp{ik{2\pi\over N}}I_N$.
$I_N$ is the $N\times N$ unit matrix. So with $k$ integer, the determinant  is one.
What is not invariant is the periodicity of fields in representations
with non-zero $N$-allity, like quark fields. So we will discard them for the moment.

Now the Wilson line, Eq.~(\ref{wilsonline}), under such a transformation is multiplied by
the Z$(N)$ phase factor $\exp{ik{2\pi\over N}}$.
So the probability to find the system with a given value for the Wilson line:
\begin{equation}
E(\widetilde P)\sim\int DA \delta(\widetilde P-\overline{ P(A_0)}~)\exp{-S(A)}
\label{effaction}
\end{equation}
has the same value in $\widetilde P$ as in $\exp{ik{2\pi\over N}}\widetilde P$, because
the measure stays the same, so does the action, only the argument of the delta function
will change. This is a most useful manifestation of Z$(N)$ symmetry. It is the analogue of the $Z$ symmetry we met in the case of QED.

Note that the Wilson line is a scalar quantity in every point $\vec x$.

So it bears resemblance to a Z$(N)$ spin variable $z(\vec x)$ defined on
a three dimensional lattice. If we endow this Z$(N)$ spin system with a
nearest neighbour
Z$(N)$ invariant action, we have a system that has a transition point
where the spin system changes from disordered into ordered behaviour.

There is now the hypothesis~\cite{mclerran} that the transition of this spin system and
that of the Yang--Mills system are in the same universality class. This is interesting
because it relates critical behaviour of a rather simple system to that of our Yang--Mills
system.

For $N=2$ and 3 this spin system is unique in that one can write down
only one action for this spin system per link:
\begin{equation}
S_{N=2,3}=\beta(z+z^*)\,.
\end{equation}
Here $z$ is a shorthand for the product of the two spin variables at the end points.

Indeed the transition is second (first) order for $N=2(3)$, and many studies have found that
critical behaviour is identical~\cite{defor}.

However for $N=4$ the spin system is not unique. Let us parametrize  the action per link
like:
\begin{equation}
S_{N=4}=\beta\big((z+z^*)+xz^2\big)\,.
\label{four}
\end{equation}

\begin {figure}[htb]
\centerline{%
\epsfig{file=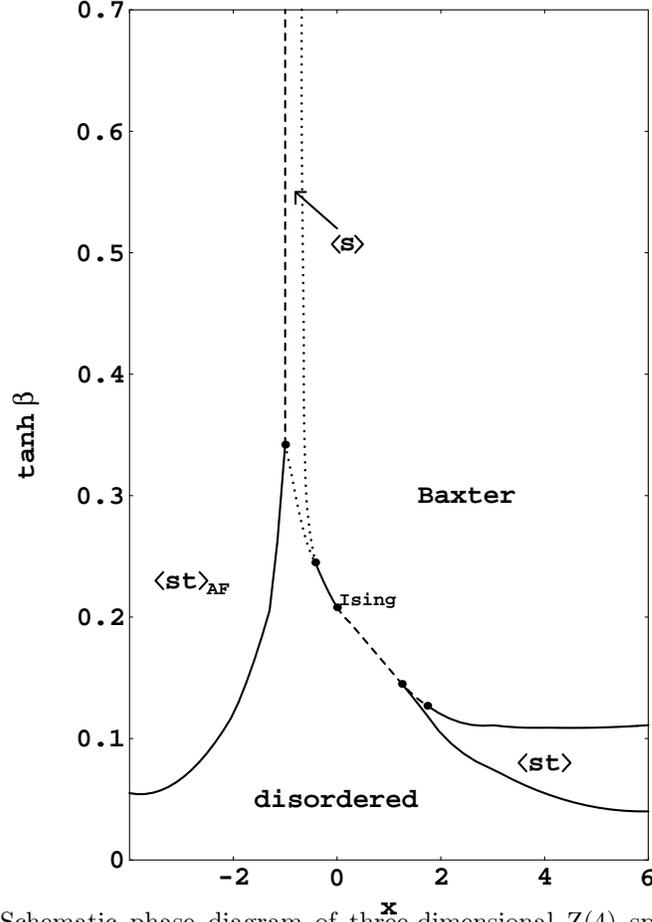,width=8.5cm}}
    \caption{%
        \label {fig:phase-diagram}
        Schematic phase diagram of three-dimensional Z(4) spin
        model~ from Eq.~(\ref{four}) on a simple
        cubic lattice, taken from
        Ref.~\protect\cite{ditzian}, where it was extracted from series
        analysis and Monte Carlo data.  Dashed and solid lines indicate
        first and second-order transitions respectively.  Dotted lines
        indicate cases where the nature of the transition has not been
        unambiguously determined.  The phases are
        labeled disordered ($\langle z\rangle=\langle z^2\rangle=0$ );
         Baxter (ferromagnetic
         with $\langle z\rangle$, $\langle z^2\rangle$ both non-zero);
         ``$\langle st\rangle $'' (where
        $\langle z^2\rangle$ is ferromagnetically ordered but
        $\langle z\rangle=0$).}
        \end {figure}

In Fig.~\ref{fig:phase-diagram}~\cite{ditzian} the phase diagram of this
 theory is plotted. Only positive couplings are of interest to us.

The VEV of the z spin corresponds to the VEV of the Wilson line $P(A_0)$ \cite{pisarski},
the VEV of  $z^2$ to that of   $P(A_0)^2)$. The region in between the two second order
transition lines corresponds to the subgroup Z(2) already broken, but not yet Z(4). That
would imply two Debye masses (not very natural from the plasma point of view), one
corresponding to $P(A_0)$ and still zero in that region. Another
corresponding to $P(A_0)^2$ and already non zero in that region.

But Nature has decided differently: in the gauge system the transition is first
 order~[10,7], and, from the phase diagram, that corresponds to both
order parameters jumping at the same time. This is what has been confirmed~\cite{teper4} in gauge theory within errors.

 Recent data~\cite{teper8} show that the first order transition becomes stronger with
increasing N. This is consistent with the idea that quasi-particles govern the behaviour
of the plasma from very high $T$ down to just
above the critical temperature.
\subsection{Electric flux and the spatial 't Hooft loop}
The phenomenon of deconfinement involves the breaking of the electric
flux tubes, and the appearance of quasi-particles, the gluons. This reflects itself in the
change in the force law between test charges, discussed above.

How does it manifest itself in other measurable quantities?  A natural
candidate is the spatial loop that measures the electric flux, as we discussed
in the first section.
This loop is formed by a closed magnetic flux line, the 't Hooft loop~\cite{thooft}.

What will perspire~\cite{kovnersteph,giovannakorthals} is  that the behaviour of
this loop
in the deconfined phase is again quite different from that in the confined phase.  The
quantitative behaviour of the loop at very high $T$ can be computed along the same lines as
in Section~\ref{sec:debyeel}. This is what we will do below.

We start with a definition of the loop as a magnetic flux loop, \ie as a gauge
transformation $\exp{(i\omega_{\rm L}(\vec x)Y_k)}$ with a discontinuity $\exp{ik{2\pi\over N}}$
when going around the loop.
Here $Y_k=\mbox{diag}~{1\over N}(N-k,N-k,...,N-k,-k,...-k)$ with $N-k$ entries $-k$ and $k$ entries $N-k$, so
that it generates the center group
element:
\begin{equation}
\exp{\left(i{2\pi} Y_k\right)}=\exp{-ik{2\pi\over N}}=z_k^*.
\end{equation}
$\omega_{\rm L}(\vec x)$ is  half the solid angle defined by the loop.

In the Hilbert space this operator reads:
\begin{equation}
\tilde V_k(L)=\exp{i\int d\vec x{1\over g} {\rm Tr}\vec E(\vec x)\cdot\vec D\omega_{\rm L}(\vec
x){Y_k\over {2}}}.
\label{thooftloopone}
\end{equation}

A representation which has the same effect in the physical Hilbert space
is:
\begin{equation}
 V_k(L)=\exp{i{2\pi\over g}\int\limits_{S(L)} d\vec S {\rm Tr}\vec E(\vec x)Y_k}.
\label{thooftlooptwo}
\end{equation}
Using the canonical commutation relations you  can check that  $V_k(L)$ and $\tilde
V_k(L)$ have the same effect on physical states. That is, they multiply
Wilson loops  with the center group factor if the latter intersects with $S(L)$. More precisely, if  the Wilsonloop $W_l(L)$ is in a representation with N-allity $l$, i.e. built from $l$ mod $N$ fundamental quark representations, then the 't Hooft commutation relation reads:
\begin{equation}
V_k(L)W_l(L')V_k(L)^{\dagger}=\exp{\{in(L,L'){2\pi\over N}\}}W_l(L').
\label{thooftcommute}
\end{equation}

The integer $n(L,L')$ equals $0$ or $1$, depending on how many times (mod 2) $L'$ intersects the surface
spanned by $L$. The reader can find more detail in ref.~\cite{kovneraltes}.

Alternatively you can {\it verbatim} follow the proof of the equality given in section {\ref{sec:debyeel}.
So again, eq.~(\ref{thooftlooptwo}) is the dual Stokes version of Eq.~(\ref{thooftloopone}).

Are the matrices $Y_k$ unique? The answer is that they can be replaced by $\Omega Y_k\Omega^{\dagger}$, with $\Omega$ a regular gauge transformation, without changing the effect of the flux operator in the physical subspace\footnote{This reflects the fact that the flux operator transforms like an adjoint under a gauge transformation.}. But that is all the freedom one has. Other choices, like generators of subgroups of $SU(N)$, lead to gauge non-invariant flux operators. 

\subsection{Quasi-particle calculation of electric loop average}
In the confined phase the particles are colour neutral so will at most contribute a
perimeter law.
At least, that is what one would infer from the analogy with the  QED flux loop in the previous section. However, for QCD we have to distinguish glueballs and bound states of quarks. A glueball is understood to be a closed fundamental
electric flux loop. So, once close enough to the loop it will send an fundamental flux through the surface, multiplying the k-loop with a centergroup factor $z_1^k$. But since it is closed, the flux has to come back through the surface giving a factor $z_1*$. These phases interfere, and thus the total effect is as if there were no charge. Only at the perimeter there is a net effect.

On the other hand a meson ( a $q\bar q$ pair) close to the loop {\it{has}} an effect: it multiplies the loop with a center group  factor $z_1^k$ or its complex conjugate, depending on its orientation. Now the phases do not interfere, and
a surface effect arises.

   Let us now  turn to the deconfined phase and repeat the quasi-particle argument for the area law:
\begin{equation}
\langle V_k(L)\rangle=\exp{-\rho_k(T)A(L)}\,.
\label{arealaw}
\end{equation}

The gluons are in the adjoint representation, so their  $Y_k$ charge follows from the
differences of the diagonal elements of $Y_k$.  That means there are $2k(N-k)$ gluons with charge
$\pm g$,
and the remaining gluons have charge $0$.

Again, the total flux of a gluon inside the slab of thickness $l_{\rm D}=m_{\rm E}^{-1}$ on both sides
of the loop
as seen by the loop is ${1\over 2}g$. The other half is lost on the loop.
So the contribution of a fixed gluon species with non-zero charge is $-1$. As we suppose
the gluons to be independent, the probability distribution for all the species inside the
slab will factorize into single species distributions $P(l)$, $l$ the number of gluons of
that species inside the slab. Only the $2k(N-k)$ species with non-zero flux will
contribute. Because of the factorization:
\begin{equation}
\langle V_k(L)\rangle=\left(\sum_lP(l)(-1)^l\right)^{2k(N-k)}
\end{equation}
and with the Poisson distribution for $P(l)$ we get $\sum_lP(l)(-1)^l=\exp{-2\bar l}$,
with $\bar l=n(T)l_{\rm D}A(L)$ the average number of the gluon species in the slab.
It follows from Eq.~(\ref{arealaw}) that the tension equals:
\begin{equation}
\rho_k=8l_{\rm D}n(T)k(N-k)\,.
\label{thoofttension}
\end{equation}

It is the dependence on the strength $k$ of the loop, which is typical for the
quasi-particle picture. We have checked in perturbation theory that  deviations of this
behaviour start to develop only to three loop order. Just above the transition we have no
reason to trust the loop expansion.
If the strong first order transition found at $N\ge 6$ really implies a
quasi-particle picture one should simulate the loop just above the transition for its
$k$-dependence. Paraphrasing ref.~\cite{teper8}, such a behaviour just above $T_c$ would suggest that the upper limit to the interface tension $\tau_{c,d}$ will scale like $N$, in accordance with a strong first order transition! To understand this argument, consider the complex plane with the possible phase of the Wilson line $P=\exp{ik{2\pi\over N}}$ on the unit circle. In the deconfined phase the effective potential
has degenerate minima in the Z(N) vacua. At $T=T_c$ one has another degenerate minimum in $P=0$, the confined phase, which in coexistence with any of the Z(N) phases, will  have the same tension $\tau_{c,d}$ because of the obvious symmetry in the plane.  A region of space with $P=1$
is separated from a region where $P=\exp{ik{2\pi\over N}}$ by a wall given by $\rho_k$, for which the tension $\sim N$ from eq.(\ref{thoofttension}). If just above the transition $T_c$ the deconfined phase with $P=0$ starts to form in between the two Z(N) phases  (so-called ``wetting'') we have $2\tau_{c,d}\le \rho_k$.  If wettting is to be true for all Z(N) interfaces the upper limit 
 follows. If we know the Z(N) spin model, that falls in the same universality class as our gauge theory, we can check these statements quantitatively\footnote{It is known that in the N state Potts model $\rho_k=\rho=\tau_{c,d}$ at the first order phase transition.}.

\section{Magnetic screening mass and spatial Wilson loop}\label{sec:debyemag}

Not only the force law between heavy electric charges like the
heavy quark, but also the force between heavy magnetic charges tells us about the medium.
The original idea of 't Hooft and Mandelstam was that of a dual superconductor, with the
electric Cooper pairs replaced by some form of magnetic condensate.  This condensate would
be expected to screen the colour-magnetic field.

In Section~\ref{sec:debyeel} we constructed an operator $V_k(L)$ creating
a magnetic flux of strength $\exp{ik{2\pi\over N}}$, Eq.~(\ref{thooftloopone}).
This loop was space like.

To get the  monopole anti-monopole pair at points $(0,r)$ we have the vortex end at $0$
and  $r$ on the positive $z$-axis. The vortex is given by a gauge transformation $V_k(\vec
x)$ which is discontinuous modulo a center group element $\exp{ik{2\pi\over N}}$ when
going around the vortex. The vortex is like the Dirac string in QED.  It is unobservable
by scattering with
particles in the adjoint representation, as long as it has center group strength. It reads, up to a regular gauge transform:

\begin{equation}
V_k=\exp{i\vec E.\vec D(A)v_k(x,y)}
\label{magnvortex}
\end{equation}
\noindent with $v_k(x,y)=\arctan({y\over x})Y_k$. When encircling the  point
$x=y=0$ the  gauge transformation $\exp{iv_k(x,y)}$ picks up a factor
$\exp{i2\pi Y_k}=z_k$. This gauge transformation remains, by definition, unchanged
along the z-direction and will be denoted by $V_k(r)$. We say that $V_k(r)$ creates a
vortex or ``Z($N$) Dirac string'' of length $r$. That means, a Wilson loop $W$ in the
fundamental representation that encircles the vortex will pick up the $z_k$ factor:
\begin{equation}
V_kWV^{\dagger}_k=z_kW\,.
\label{wilsonlooptwist}
\end{equation}

 Any Wilson loop with non-zero $N$-allity
$l$ will pick up a factor $(z_k)^l$. But Z($N$) neutral loops will not sense
the Z($N$) Dirac string, hence the name.

On the lattice the Hamiltonian operator will have magnetic plaquette operators.
These will pick up the $z_k$ factor and it is not hard to see that
the Gibbs trace can be worked into a path integral along the usual lines, and on the
lattice the latter takes the form:
\begin{equation}
\exp{{-{F_{\rm M}(r)\over T}}}={\int DA\exp{-S_{(k)}(A)}\over\int DA\exp{-S(A)}}.
\label{latticemagnetic}
\end{equation}

The action $S_{(k)}$ is the usual lattice action, except for those plaquettes pierced by
the Dirac string. Those plaquettes are multiplied by a factor $\exp{ik{2\pi\over N}}$, as
in Fig.~\ref{fig:chain}. This string is repeated at every time slice between $\tau=0$ and
$\tau=1/T$.

\begin{figure}[htb]
\centerline{%
\epsfig{file=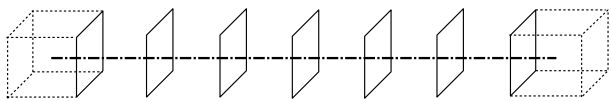,width=8cm}}
\caption{\label{fig:chain}Monopole antimonopole pair induced by
twisting the plaquettes pierced by the Dirac string.}
\end{figure}

Screening is expected  in both confined and deconfined phases:
\begin{equation}
F_{\rm M}(r)=F_{{\rm M}0}-c_{\rm M}{\exp{-m_{\rm M}r}\over r}\,.
\end{equation}
All parameters are function of $T$.
 In the cold phase the screening is a consequence of the electric flux confinement. This
is natural because the ground state contains a condensate of ``magnetic Cooper pairs'',
according to the dual superconductor analogy. It is a screening mechanism whose details
are not understood. We dropped for notational reason the dependence on the strength  $k$
of the monopole in the coefficient $c_{\rm M}$.

In the hot phase there are indications from spatial Wilson loop simulations  that there is
additional thermal screening from magnetic quasi-particles, as discussed in Section~\ref{sec:model}.
The magnetic mass does  not depend  on the strength $k$ of the source, just like
the Debye mass does not, as discussed in the previous section. What we will
show below is that the magnetic mass is a $0^{++}_+$ state in the fictitious Hamiltonian $\hat H$.

Analogous to  the Wilson line correlator we consider the Hamiltonian $\hat H$ in the
fictitious system of $(x,y,\tau)$ variables. We search  the  operator $V_k$ acting on the
Hilbert space of physical states  of this Hamiltonian, that reproduces the path integral
Eq.~(\ref{latticemagnetic})~\footnote{We use the same notation
as for the vortex operator in $(x,y,z)$ space as there is no risk for confusion.} . So
$V_k$ should create a vortex in the $(x,y)$ plane at
every time slice $\tau$ and the Hamiltonian $\hat H$  should propagate every one of these
vortices in the z-direction over a distance $r$. So $V_k$ is the 't~Hooft vortex operator
discussed around Eq.~(\ref{magnvortex}):

\begin{equation}
V_k=\exp{i\int\limits_{x,y,\tau}{\rm Tr}\vec E.\vec D(A)v_k(x,y)}.
\end{equation}

\noindent with $v_k(x,y)=\arctan({y\over x})Y_k$.

Both under parity (remember: only $y\rightarrow -y$!) and charge conjugation the vortex
$V_k$ transforms into $V_k^{\dagger}$. R parity leaves it invariant.  Its spin $J$ equals $0$, despite the appearance
of the rotated
singularity line. On physical states the location of the singularity does not matter.
Hence the operator  Im$V_k$ excites spin zero states with $R=1$ and $P=C=-1$, and Re$V_k$
excites scalars with $R=P=C=1$. Which combination is the one we are after?
In contrast to the Debye mass, we  continue for $T$ below $T_c$ to measure the magnetic mass. At $T=0$ it becomes a {\it three dimensional} scalar. And with 3d rotational symmetry restored  $R$ and $P$ parity are related by a rotation. Therefore, on a scalar state $P=R$. Thus Im$V_k$ is excluded, and the magnetic mass is just that of the $0^{++}_+$\footnote{This fact was overlooked in earlier work, hep-ph/0308229, published in Vol. 40 of
Erice International School of Physics, ``From Quarks and Gluons to Quantumgravity, pg 72-73 and in Act. Phys.Pol. B, 2003, 12 : 5825.}. 

Perturbation theory is not reliable  and we need lattice
simulations~\cite{rebbi}. Up to now these simulations are four dimensional and have
limited accuracy. They need to be repeated, also in dimensionally reduced form. Once they
reproduce
the mass levels of the fictitious Hamiltonian with sufficient accuracy,
we can use them with more confidence for determining the tension of the space-like 't
Hooft loops.

At high $T$ where reduced calculations are valid we expect to find the mass level of the
$0^{++}$ of the reduced Hamiltonian.
 From Teper's work~\cite{teper3d}   the lowest  $0^{++}$ mass in units of the string
tension  for SU($N$) ($N\ge 3$) gives in a large $N$ expansion (see his table 34):
\begin{equation}
{m^{++}\over{\sqrt\sigma}}=4.065(55)(1+0.634(20)/N^2 +O(1/N^4))\,.
\label{dilute}
\end{equation}
For $N=2$ one finds~\cite{teper3d} for this ratio $4.718(43)$, for $N=3$ it is $4.329(41)$.

\subsection{The spatial Wilson loop}\label{wilson}

The spatial Wilson loop is of interest because it monitors the magnetic activity in the
plasma. At zero temperature it obeys an area law identical to that of
time-like loop, controlled by the zero temperature string tension.

The tension stays constant throughout the confined phase, and starts to rise about the
critical temperature, indicating a new source of magnetic flux activity.

Let us begin with some basics: a representation built from $k$ fundamental representations
is said to have $N$-allity $k$. A center group transformation  $\exp{i{2\pi\over N}}$ is mapped
into
$\exp{ik{2\pi\over N}}$ in such a representation.
Write a Wilson loop formed with such a representation as $W_k(L)$.
Its average will then give an area law with tension $\sigma_k$.
 It is generally accepted that the Wilsonloop tension does not depend on the chosen representation, only on the N-allity $k$.

At high $T$ one can integrate out the hard modes, as they do not determine the
string tension. One can also integrate out the electrostatic modes, and wind up with a
path integral controlled by the magnetostatic action:
\begin{equation}
\exp{-\sigma_k(T)A(L)}={\int  D\vec A W_k(L)\exp{-S_{\rm M}(A)}\over \int
D\vec A)\exp{-S_{\rm M}(A)}}\,.
\label{eq:wloopmagnetic}
\end{equation}
The hard and electrostatic free energies $f_{\rm h} ~\mbox{and}~ f_{\rm E}$ drop out in the ratio.

The only dimensionful scale in the magnetostatic action is $g_{\rm M}^2$. So
the tension, having dimension $\mbox{(mass)}^2$, can be written as:
\begin{equation}
\sigma_k(T)=c_kg_{\rm M}^4\left(1+O(g^3)\right).
\label{eq:wtensionmagnetic}
\end{equation}

So the dominant contribution to the tension is entirely from the magnetostatic sector.
In  Fig.~\ref{fig:string} you see a fit of the tension data to this parametric
expression for SU(3).

\begin{figure}[htb]
\centerline{
 \epsfig{file=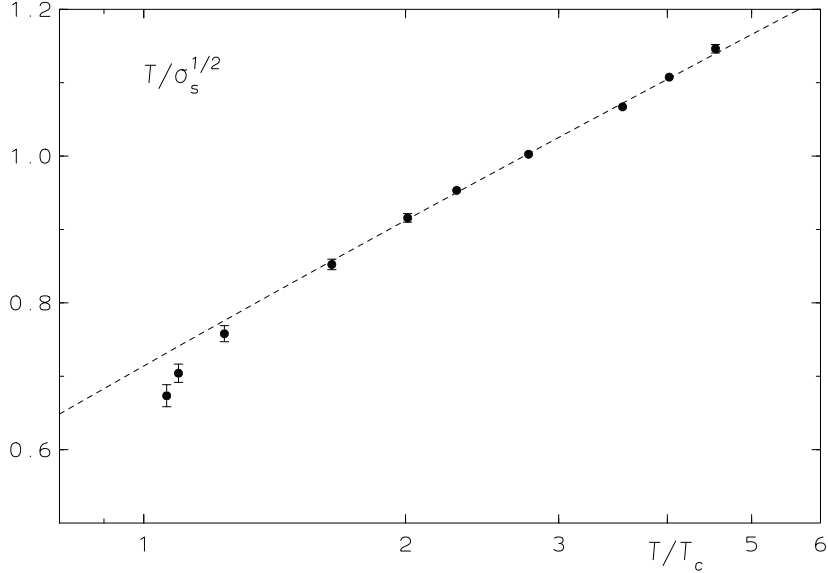,width=9cm, angle=-90}}
\caption{The temperature over the square root of the spatial string tension $\sigma_s$
 {\it versus} $T/T_{\rm c}$ for SU(3). The dashed line shows a fit
according to a two loop scaling formula for the coupling, see text below
Eq.~(\ref{eq:wtensionmagnetic}). From Ref.\cite{karschtension}. }
\label{fig:string}
\end{figure}

The authors took for the magnetic coupling $g_{\rm M}^2=g_{\rm E}^2$, so neglected renormalization
effects of the scale $gT$, which are a few percent at $T=2T_{\rm c}$ (see eq.(\ref{elmag2})) On the other hand they
included two loop renormalization effects in the coupling. Dropping the latter effects (as we argued before), and taking into account the renormalization of $g_M$ as in eq.(\ref{elmag2}) the data points in fig. (\ref{fig:string}) constrain through a $\chi^2$ analysis the values of $c_{k=1}$ and $T_{\rm c}/{\mit\Lambda}_{\overline{MS}}$. We show this in fig.(\ref{fig:cl}).

\begin{figure}[htb]
\centerline{
 \psfig{file=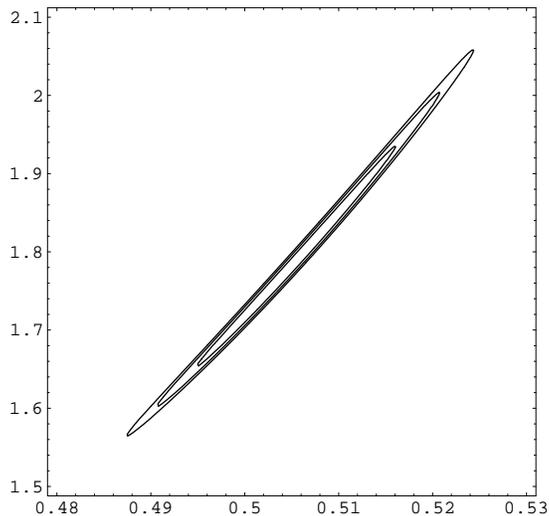,width=9cm, angle=-90}}
\caption{Confidence limits (90 resp. 68 \% for outer, resp inner  ellipse) for $c_{k=1}$ (abscis) and $T_{\rm c}/{\mit\Lambda}_{\overline{MS}}$ (ordinate) for SU(3) from the data down to $2T_c$ in fig.(\ref{fig:string}). From Ref.\cite{pierre}. }
\label{fig:cl}
\end{figure}
The value of  $c_{k=1}=0.5530(20)$ follows from ref.\cite{teper3d} and is not compatible with the confidence limits. The central values
for the ratios $T_{\rm c}/{\mit\Lambda}_{\overline{MS}}$ are off the quoted values \cite{pierre}. Taking only data points above $3T_c$ does not improve the situation. 

It follows that 4D  data for higher than $4T_c$ are needed in order to see the
3D physics. For more detail the reader should consult ref.~~\cite{pierre}.

The spatial Wilson loop measures in a sense to be specified later the magnetic flux in the
system. As we will see in section (\ref{sec:model}) the 3D loops are well described by a magnetic quasi-particle picture. So this picture may be valid down to  a few times the critical temperature. 

On the other hand the tension is flat from $T=0$ to $T=T_{\rm c}$ within errors, according to the data.  In all of the
confined phase the magnetic activity does not
change.

\section{A simple model}\label{sec:model}

 In close analogy with the 't Hooft loop one can do a quasi-particle 
 calculation for the
k-tension of the Wilson loop. But what are the magnetic quasi-particles?

We assume here that the magnetic screening length $l_M$ defined in section
(\ref{sec:debyemag}) defines three dimensional non-perturbative lumps,
called magnetic quasi-particles. They are constituted by the magnetic gluons.

Such a lump at rest will have a screened colour magnetic field.

Note that the magnetic screening length defines a volume in which many
elementary quanta, the magnetic gluons, are present, just like in the case of the Debye screening length.
 The difference is that the Debye sphere, though containing many elementary gluons , is not a bound state of the particles it contains.

We are making the simplest possible assumptions:
\begin{itemize}
\item
\hspace{-0.2cm}The magnetic quasi-particles have  a screening length $l_{\rm M}\sim g^2T$ much smaller than
their average distance, i.e. they form a dilute gas.
\item
\hspace{-0.2cm}The magnetic quasi-particles are in the adjoint representation of the magnetic  $SU(N)$ group.
\end{itemize}

The global magnetic group as introduced in ref.\cite{olive} should be either $SU(N)$
(in case there is only Z(N) neutral matter) or $SU(N)/Z(N)$ (in case Z(N) charged matter like quarks couples). So our adjoint representation covers both.   

We should have, like for the 't Hooft loop, a magnetic flux representation
for the $k$-Wilson loop. One can argue, starting from existing results~\cite{diakonov},  that for the irrep constructed by fully antisymmetrizing k quark representations the  average of the corresponding Wilson loop can be computed from:

\begin{equation}
\left\langle W_k(L)\right\rangle=\left\langle \exp{\{ig\int d\vec S\cdot{\rm Tr}\vec
B Y_k\}}\right\rangle\,,
\label{wilsonflux}
\end{equation}

\noindent where, as before, $Y_k=\mbox{diag}~{1\over N}(N-k,N-k,...,N-k,-k,...-k)$
with $N-k$ entries $-k$
and $k$ entries $N-k$, so that it generates the center group
element:
\begin{equation}
\exp{\left(i2\pi Y_k\right)}=\exp{-ik{2\pi\over N}}=z_k^*.
\label{yoncemore}
\end{equation}
For details on eq.(\ref{wilsonflux}) the reader should consult Appendic B and C.

The $Y_k$-charge of a magnetic quasi-particle is $\pm {2\pi\over g}$
with the same multiplicity $2k(N-k)$ as for the electric quasi-particles. It contributes $-1$ to the Wilson loop
Eq.~(\ref{wilsonflux}) because only one-half of its flux goes through the
loop.

With our assumptions we can now, in precise analogy with the calculation
of the 't Hooft loop, understand why the $k$-tension scales like:
\begin{equation}
\sigma_k=c k(N-k) l_{\rm M} n_{\rm M}(1+O(\delta))\,.
\label{kscaling}
\end{equation}
\noindent with $\delta$ the diluteness parameter $l_M^3n_{\rm M}$. This parameter, as argued at the end of the next subsection, turns out to be about $0.06$.
As in the gluon case, $n_{\rm M}$ is the density of one quasi-particle species.
The constant c represents the effects of using the Poisson distribution for a single quasi-particle species and some elementary geometry, as in
section (\ref{sec:debyeel}), eq.(\ref{thoofttension})). So it is a number, in which colour dependence is absent. It is calculated in Appendix A to be $c=6.57128...$.  We will come back to this point in the next subsection. 

The simplest test of this k-scaling is to consider the ratios of tensions in which  everything
drops out, except the k-dependence. The tensions, as far as their leading term is concerned, can be simulated in 3 dimensions.

 Mathematically the degeneracy of the number of charged particles in the adjoint representation has the same k dependence as the quadratic Casimir
operator of the totally antisymmetric representation built from $k<N$ quarks. This quadratic Casimir equals $C_Fk(N-k)/(N-1)$, with $C_F=(N^2-1)/2N$ the the Casimir of the fundamental representation. We will refer to our predicted k-dependence as Casimir scaling.

\subsection{Comparison to lattice results}
Our prediction is for  the totally antisymmetric representation with N-allity $k$.

At the time the model was proposed, there were only tensions known for SU(2) and SU(3) groups.  There our prediction for the ratios is already implied by charge conjugation.

 Since the last three years lattice data on string tensions for $N\ge 4$ have been taken.
The ratios found~\cite{luciniteper} for the totally antisymmetric irreps are close --- within a percent
for the central value --- :
\begin{eqnarray*}
{\rm SU}(4):\sigma_2/\sigma_1&=&1.3548\pm 0.0064~~~~\hbox{ Casimir} :1.3333  \\
{\rm SU}(6):\sigma_2/\sigma_1&=&1.6160 \pm 0.0086~~~~\hbox{ Casimir} :1.6000 \\
            \sigma_3/\sigma_1&=&1.808\pm 0.025~~~~~~~\hbox{ Casimir} :1.8000
\end{eqnarray*}

The results are that precise, that you see a two standard deviation, except
 for the second
 ratio of $SU(6)$. This deviation is natural, since the diluteness of the
magnetic quasi-particles is small, on the order of a couple of percent, as we will explain at the end of this subsection. So we expect corrections on that order to our ratios. 

There is a less precise determination of the ratio
 $\sigma_2/\sigma_1=1.52\pm0.15$ in
$SU(5)$~\cite{meyer}. But the central value is within $1$ to $2\%$ of the
predicted value $3/2$.

The $SU(8)$ ratios are known on a rather course lattice~\cite{meyer} and using a different algorithm:

\begin{eqnarray*}
\sigma_2/\sigma_1 & = & 1.692(29)~~~~\hbox{ Casimir} : 1.714\\
\sigma_3/\sigma_1 & = & 2.160(64)~~~~\hbox{ Casimir} : 2.143\\
\sigma_4/\sigma_1 & = & 2.26(12)~~~~~\hbox{ Casimir} : 2.286\\
\end{eqnarray*}

In conclusion: the seven measured  ratios are consistent with the quasi-particles being
independent, as in a dilute gas. The number of quasi-particle species contributing
to the k-tension is $2k(N-k)$.

A posteriori, one can understand the diluteness being small. Take our 
formula for the tension, and multiply it with $l_M^2$ to get the diluteness:
\begin{equation}
l^3_Mn_M\le l_M^2\sigma_k
\label{smalldilute}
\end{equation}

The diluteness  has no parametric reason to be small. But from lattice data~\cite{teper3d} one finds typically $0.05$~\footnote{This small ratio is naturally explained  in flux tube models of the spectrum. The electric analogue of this ratio is $l_D^3n\sim 1/g^3$, and is large precisely for parametric reasons.}, as we discussed in section (\ref{sec:debyemag}).

\subsection{Partial pressure, density and N dependence of the tension}

There is an important  question left. It concerns the behaviour of the tension
for large N. In our expression for the tension of the 
Wilson loop we had a constant $c$, independent of the number of colours. The magnetic screening $l_M$ is $O(1)$ as argued above. Now, if we would take the density of one species to be:
\begin{equation}
n_M=d(N)(g^2NT)^3,
\end{equation}
\noindent then the large N behaviour of $\sigma_1$ would be determined by the product $d(N)N$.
If we would take naively $d(N)=O(1)$, that is the density of one species to be $O(1)$,  
  this would contradict the result from all orders in perturbation  theory, that the Wilson loop tension is $O(1)$ for large $N$! 
So we are forced to  insist on a one species density
of order ${1\over N}$, i.e. $d(N)={d\over N}$.

This dependence on N, combined with the idea that the magnetic quasi-particles 
are free, has obviously consequences for the pressure of the plasma discussed 
in  section~\ref{sec:eff}. 

The effects
from the magnetostatic action showed up at $O(g^6)$. This contribution as discussed extensively there, contains the Stefan-Boltzmann factor $\sim (N^2-1)T^4$,
multiplied with $(g^2N)^3$( and eventual non-leading terms of order ${1\over {N^2}}$). 

If we believe - as  seems reasonable from the  lattice data discussed before - that the magnetic quasi-particles are independent, then they should deliver a partial pressure proportional to $N^2-1$ and the density of a given fixed species $n_M={d\over N}{(g^2NT)^3\over {\pi^2}}$. We put the 't Hooft coupling in front of the temperature as it should in the 3d theory.  

  Let the free energy per quasi-particle be $f(N) T$. So the partial free energy from all the quasi-particles will be $(N^2-1){d\over N}{(g^2NT)^3\over {\pi^2}}f(N)T$, with the product ${d\over N} f(N)$ of O(1).

  So this implies $f(N)\sim N$ for the free energy per magnetic quasi-particle.
 In other words the colour degrees of freedom of the lumps do show up in the free energy. It means the lumps have an energy $N\epsilon-\delta$, where $\delta$ is the binding energy growing less fast than $N$. 
Any simulation of the partial free energy would then tell us the value of $df(N)$.

\section{Conclusions}

In these lectures we concentrated  
on the quasi-particle picture of the electric and  magnetic sector. The lattice data are
consistent with the predictions within  a few percent, the typical order of magnitude of
their diluteness.

 The magnetic quasi-particle model is certainly a viable idea, and it deserves study in the Minkowski description. After all, this is where plasmas live.  It is perhaps not superfluous to remind the reader that magnetic quasi-particles have a density that is suppressed by $(\log({T\over {\Lambda}}))^3$. At asymptotic temperatures we have the usual Stefan-Boltzman gas.

Of course, lattice simulations stay of invaluable support and are an indispensible testing ground.
A most probably incomplete list of suggested simulations follows.
\begin{itemize}
\item{
Determination of the partial three-dimensional pressure from lattice simulations~\cite{york}. It would further constrain the model.}

\item{At large $N$ the strong first order results reported by Teper~\cite{teper8} suggest a quasi-particle picture
down to the critical temperature and can be put into evidence  by simulating  at $T^+_c$ spatial Wilson and 't
Hooft loops. The simulation of the latter is now getting in a new stage~\cite{defor}, where we start to
learn~\cite{owe} the systematic errors from comparison with the known
screening masses. The area law of the spatial 't Hooft loop permits simulation  by multilevel methods~\cite{meyer}. From the results on the 't Hooft loop at $T_c^+$ one could infer the corresponding $Z(N)$ spin model.} 

 \item{If at large N indeed the quasi-particle picture is valid down to $T_c$, one would expect the topological susceptibility to drop to zero at $T_c$. This tendency is confirmed in a recent paper~\cite{tepertopol} and by the Pisa group~\cite{luigi}. Here one needs to distiguish exponential decay
due to screening, valid for all $N$. }

\end{itemize}

On  the theory side there are many questions that need more scrutiny. 
\begin{itemize}
\item{Our model gives  results for the AS irrep, depending on the N-ality. Independence of the representation
 is vindicated  by lattice methods~\cite{luigisu3} for SU93). Taking this independence for granted there are arguments~\cite{armoni}
 that the large N  limit of $\sigma_k$ admits only corrections from powers of $1/N^2$. Our leading order result is inconsistent with this claim, but it may be that the series in the diluteness will undo the $1/N$ corrections. Of foremost importance is the apparent accuracy of our result, better than $1/N^2$ for $N\le 4$.}
\item{The use of the flux formula, as advocated in ref.~\cite{diakonov}, can be justified in a more specific way. In appendix C we argue why eq. (\ref{wilsonflux}) is  the one that applies to the AS irrep.}
\item{An amusing speculation can put the Wilson loop and the 't Hooft loop
on an equal footing. Suppose that the magnetic group is operating at scales $g^2T$ as a local gauge group. Then we could define the Wilson loop as a magnetic gauge transform with discontinuity in $Z(N)$. The area law would then trivially depend only on N-allity. Also the magnetic flux formula for the loop, eq.~(\ref{wilsonflux}), would be implied.}
\item{How do magnetic quasi-particles behave at high density, especially in the CFL phase with a VEV for the glue? Does one get bound state formation?} 
\end{itemize}

It is amusing to make an estimate of the number of magnetic quasi-particles
in the typical  lattice volumes so far considered. This can be easily extracted from the values of $l_M$ in terms of the lattice length $a$ ($\ge 1$), and the lattice size $L$ in terms of $a$ ($\sim 20$). The diluteness of one species, eq. (\ref{smalldilute}), gives a number on the order of a few hundred for one species and say $N\le 4$.

How can one view the mechanism of the transition in terms of the quasi-particles? First, at temperatures slightly above $T_c$, one is confronted with a thermal de Broglie wave length $T$ of the same order as the interparticle distance $g^2T$ and the Bose-Einstein character of the quasi-particles starts coming into play. So the statistical independence is no longer a viable approximation. 
One expects the magnetic quasi-particles to condense in the groundstate below $T_c$. There they give rise to a non-zero string tension through electric flux tube formation.  The simplest assumption--reasonable because also in 4D the diluteness as given by the  ratio $l_M^2\sigma$ is small, about $0.09$-- is that one has a   condensate of our dilute Bose gas . To leading order in the density
  expansion, there are now however
two and higher particle correlations. They will give deviations from the single particle multiplicity, that gave us in the plasma phase the Casimir scaling.
 
 At the same time their effect on the spatial Wilson loops is constant in $T\le T_c$ as we know from the simulations. This is an  important clue to understand. The tension of the time-like Wilson loops is related to the fraction of superfluid quasi-particles and is known to diminish as the temperature rises. So one would expect this tension to decrease when approaching  $T_c$ from below. This is qualitatively the case in lattice simulations.  

\bigskip\medskip

I thank Pierre Giovannangeli, Ben Grypaios, Alex Kovner, Laurent Lellouch,  Rob Pisarski, David Skinner,  Mike Teper, Mikko Laine and  Urs Wenger  for  constructive and sometimes hard-needed deconstructive criticism. A Royal Society grant permitted me 
to write these notes in  the stimulating atmosphere of the Oxford Dept of Theoretical Physics. Special thanks go to Harvey Meyer for communicating  his preliminary $SU(8)$ data on Wilson loops.

\section{Appendix A}
It is
of interest to figure out how in our independent quasi-particle picture
the average of $V_k(L)$ will behave as function of its strength k. As a byproduct we treat the geometry more precisely.

A simple argument shows that it must be less than linear in k. First note
that $V_k(L)=(V_1(L))^k$. Imagine now we pull the k factors apart into an array
 of k parallel $V_1(L)$ loops separated by distances
well over twice the screening length. Then the loops get uncorrelated and the result
for the array is k times the result for $V_1(L)$. Thus we have a linear law, $k\rho_1$, when the unit strength loops are far apart, because each of them gets
disordered by the fluxes of the particles in its surrounding  slab of thickness $2l_D$. 

Now we start to collapse the array. If we keep the loops in the array just one Debye length apart, any charge in a slab will disorder the two
loops limiting the slab. This introduces correlations and the linearity 
may change.

An estimate is based on the idea that infinitesimal slabs at distance $l$, and of width $dl$, are
{\it {independently}} contributing to the average $<V_k>$. So the average is built up
from a product $\Pi_l<V_k>_{l,dl}$. 

Let us denote by $\Phi(l)$ the  flux normalized by the charge $e$ at distance $l$ from the loop, thought to be large and circular with radius $R$ and area $A$.

The average from one charge species from two slabs at distance $\pm l$ and of thickness $dl$ is:
\begin{equation}
<V_k>_{l,dl}<V_k>_{-l,dl}=\sum_{m\ge 0}P(m)\exp{\{ikm\Phi(l)\}}\sum_{n\ge 0} P(n)\exp{\{-ikn\Phi(l)\}},
\end{equation}
\noindent where we used the independence.

 Take for  $P(n)$  the Poisson distribution  to get:
\begin{equation}
<V_k>_{l,dl}<V_k>_{-l,dl}=\exp{\{-dl A n(T)2(1-\cos(2k\pi\Phi(l)))\}}
\label{dl}.  
\end{equation}
So the contribution from all slabs is:

\begin{equation}
<V_k>=\exp{\{-2An(T)\int_{0}^{\infty}dl(1-\cos(2k\pi\Phi))\}}.
\label{dlintegrated}  
\end{equation}

Note that a unit strength loop gets the same value when the gas is consisting of charges of strength $k$.

For the total flux going through our loop one  finds in terms of the 
screened potential $A_0={e\over{4\pi r}}\exp{-r/l_D}$
very simply:
\begin{equation}
\Phi(l)={2\pi l\over e} (A_0(r=l)-A_0(r=\sqrt{l^2+R^2})).
\label{classicfluxe}
\end{equation}

Plugging this into eq. (\ref{dlintegrated}) and dropping the corrections due to the size $R$ of the loop, one gets for the tension for one single species:
\begin{equation}
\rho_k(T)=2 n(T) \int_{0}^{\infty} dl(1-\cos(k\pi\exp{-l/l_D})).
\label{classicflux}
\end{equation}

For $k=1$ the integral $\int_{0}^{\infty} dx(1-\cos(\pi\exp{-x})=1.64282...$.

The derivative with $k$ of $\rho_k$ is easily computed:
 \begin{equation}
{\partial\over{\partial k}}\rho_k(T)=2{l_Dn(T)\over k}(1-\cos\pi k).
\label{logkscaling}
\end{equation}

For odd k one finds from this:
\begin{equation}
\rho_k(T)=4l_Dn(T)\log k +\rho_1(T).
\end{equation}

The even k lie just below the logarithmic curve, and have derivative zero.
 So this k-dependence is like a  logarithmic staircase.

As we noted above, the Z-ionic tension for the unit flux loop is identical to the electronic tension
 of a loop with $Z$ units of flux. So the slow rise of the tension with $Z$ flux units shows there is
binding  compared to a configuration of $Z$ distant loops with unit flux.
 
Note that for the non-Abelian k-loops there is {\it {no}} k dependence  for a single quasi-particle species, since the charge is $\pm g$, {\it{ independent}}  of $k$. The k dependence comes in through the {\it{multiplicity}} of the charged particles.

\section{Appendix B}

In this appendix we briefly indicate the group theory needed to get from a given Young tableau (defining the irreducible representation $R$) the corresponding highest weight and the value of the quadratic Casimir.In what follows we suppose a representation to be irreducible without mentioning so.                 

Let the Young tableau have $n_1$ boxes in the first row, $n_2$ in the second row, etc.. Then one can define the non-negative numbers $w_l=n_l-n_{l+1}$.

Now the highest weight matrix for the Young tableau is defined from our $Y$ matrices, defined 
above eq.(\ref{yoncemore}) by:
 \begin{equation}
H_R=\sum_{l=1}^{N-1}w_lY_l.
\label{highweight}
\end{equation}

For example, for the totally antisymmetric tableau of k boxes in one column we have $H_R=Y_k$. For the totally symmetric tableau with all k boxes 
in one row $H_R=kL_1$.
Note that the stability group of $Y_k$ (the subgroup of $SU(N)$ matrices  commuting with $Y_k$) is $SU(k)\times SU(N-k)\times U(1)$.

So the totally antisymmetric representation with k squares has a highest weight
with this stability group. All other representations with k squares have different stability groups.

We define one more diagonal $NxN$ matrix by:
 \begin{equation}
2Y\equiv 2\sum_{l=1}^{N-1}=diag (N-1,N-3,....,-N+1).
\end{equation}

The quadratic Casimir operator $C_2(N,k,\{w_l\}\equiv C_2(R)$ is defined
by summing the square of all generators $T_a$ in the representation $R$. The result is $\sum T_a^2=C_2(R) 1_{R}$, where $1_{R}$ is the unit matrix in $R$ and $C_2(R)$ is a c-number (normalization is $[T_a,T_b]=if_{abc}T_c, f_{abc}f_{bcd}=N\delta_{ad}$).
Then  the quadratic Casimir equals:
\begin{equation} 
C_2(R)={1\over 2}(TrH_R^2+2TrYH_R).
\end{equation}

The quadratic Casimir for the fundamental representation is $C_F={(N^2-1)\over{2(N-1)}}$
The Casimir for the antisymmetric representation is then
\begin{equation} 
C_2(R=AS)=C_F{k(N-k)\over{(N-1)}},
\end{equation}

\noindent and for the symmetric representation it is:
\begin{equation} 
C_2(R=SS)=C_F{k(N+k)\over{(N+1)}}.
\end{equation}

One can show that for fixed $k\le N$ the antisymmetric Casimir is the minimal one.

\section{Appendic C}
In this appendix we review the Stokes formula~\cite{diakonov} for the Wilson loop in the representation $R$. Let $\Omega$ be any gauge transformation that is periodic on the loop. Then, with $\vec\nabla=\vec\partial-ig\vec A$:
\begin{equation}
W_R(L)=\int D\Omega\exp{\{ig\int d\vec S.Tr[H_R\big(\Omega\vec B\Omega^{\dagger}-{1\over g}\vec\nabla\Omega\times \vec\nabla\Omega^{\dagger}\big)]\}}.
\label{diakonov}
\end{equation}

If $R$ is the antisymmetric representation we have from the formulae for the highest weight in Appendix B the equation in the text, eq.(\ref{wilsonflux}).
Note that this representation is the unique one with stability group  $SU(k)\times SU(N-k)\times U(1)$.

 The integration over the regular gauges has been dropped in the main text because physical states do not feel them. Physical states include monopole configurations, which is why the second term in eq. (\ref{diakonov})
is discarded in eq.(\ref{wilsonflux}).

There is a comment related to this Stokes formula (see Appendix D). It is derived under the assumption that it is regulated by the SU(N) asymmetric top~\cite{diakonov}). The question is  whether the pure Yang-Mills theory average can be
provided with  such a regulator. For $N=2$ and in three dimensions the answer
is affirmative~\cite{kovneraltes} by adding an adjoint Higgs system and letting the VEV go to zero, followed by decoupling the Higgs in the infinite mass limit. For general N the answer is analogous, but Nature realizes only a limited set of  Higgs phases. They are limited to breakings of the type where $SU(k)\times SU(N-k)\times U(1)$ is still unbroken, k integer and $\le [N/2]$~\cite{rajantie}.
That implies that the Stokes formula is only valid for those highest weights that have this symmetry, i.e. the totally antisymmetric ones with k boxes. This justifies eq. (\ref{wilsonflux}) in the main  text.

\section{Appendix D}
The derivation of eq.(\ref{diakonov}) is based on Fourier analysis on
the group $SU(N)$, and on simple properties of the quantum-mechanical SU(N) rotator. 

We are interested in the Wilson line between two points $x(s_1)$ and $x(s_2)$, the line between the two points being parametrized by $s$. The line 
is the ordered product of some irrep $R_0$ with highest weight $H_0$ of unitary matrices:
\begin{equation}
W(s_2,s_1)=P\exp{ig\int_{s_1}^{s_2} A_sds}.
\end{equation}
Here $A_s={d\vec x\over {ds}}.\vec A$ the projection of $\vec A$ on the
line.

$W(s_2,s_1)$ is a unitary matrix in the representation $R_0$ with highest weight $H_0$.

 Along the line one can write the vector potential as a pure gauge
\begin{equation}
 A_s={-1\over{ig}}U\partial_sU^{\dagger}
\end{equation}
\noindent and so one can gauge away  $A_s$ except at the end points  :
\begin{equation}
W(s_2,s_1)=D^{R_0}(U_2U_1^{\dagger}).
\label{puregauge}
\end{equation}

To get the Wilsonloop we let $s_2\rightarrow s_1$ and take the trace of eq.(\ref{puregauge}). In this form it is not hard to show that the Wilson loop is the
Fourier component of the   propagator of the SU(N) rotator with Hamiltonian ${\cal{H}}$:
\begin{equation}
\int d\Omega\langle \Omega U_2|\exp{\{i(s_2-s_1){\cal{H}}\}}|\Omega U_1\rangle.
\label{propagator}
\end{equation}
\noindent in the limit of vanishing moment of inertia $I$.
To define the Hamiltonian we start with the 
  Lagrangian $L$ of the rotator. Define the angular velocities $V_a=Tr\Omega\partial_s\Omega^{\dagger}\lambda_a$. In terms of the Cartan basis $\lambda_{ij},i\neq j=1,....,N$ and $\lambda_d, d=1,...,N-1$  it is given by:
\begin{equation}
L={I\over 2}\{\sum_{ij}V_{ij}^2+\sum_dV_d^2\}+\sum_d V_d H_d.
\label{lagrange}
\end{equation}

The highest weight $H_0$ of our representation $R_0$ is written as $H_0=\sum_dH_d\lambda_d$.
This Lagrangian gives the Hamiltonian in terms of the canonical momenta $J_a$:
\begin{equation}
{\cal{H}}={1\over{2I}}\{\sum_{ij}J_{ij}^2+\sum_d(J_d-H_d)^2\}={1\over{2I}}\{C_2-\sum_d J_d^2+\sum_d(J_d-H_d)^2\}
\label{hamilton}
\end{equation}

Inserting a set of intermediate states,  eigenstates of the Hamiltonian, labeled by $R$,and $\{m_d=J(R)_d\}$,  gives for the matrix element of the propagator $\langle U_2\Omega_2|\exp{\{-i(s_2-s_1){\cal{H}}\}}|U_1\Omega_1\rangle$:
\begin{eqnarray}
 &{}& \sum_{R,\{m_d\}}\langle U_2\Omega|R,\{m_d\}\rangle\exp{\{-i{(s_2-s_1)\over{2I}}(C_2(R)-\sum_dm_d^2)\}} \\ \nonumber
&\times&\exp{\{-i{(s_2-s_1)\over{2I}}(\sum_d (m_d-H_d)^2)\}} \langle R,\{m_d\}|U_1\Omega\rangle.
\label{intermediate}
\end{eqnarray}

Now we have, writing $m={m_d}$:
\begin{equation}
\langle U_2\Omega|R,m\rangle\langle R, m|U_1\Omega\rangle =d_RD_{m,m}^R(U_2\Omega)(U_1\Omega)^{\dagger})
\label{matrix}
\end{equation}
\noindent with $d_R$ the dimensionality of the representation.
 Let $I\rightarrow 0$ in the exponent. This forces $H_d=m_d$ and $C_2(R)=C_2(R_0)$. The other intermediate representations have higher quadratic Casimirs.

So in this limit the integral over  the matrix element, eq.(\ref{propagator}), reduces to
\begin{equation}
\int d\Omega d_{R_0} D^{R_0}_{H,H}( \Omega U_2(\Omega U_1)^{\dagger})\exp{\{i(s_2-s_1)(C_2(R_0)-{1\over 2}TrH_0^2)/2I\}}.
\label{limitmatrix}
\end{equation}
Now the representation matrices $D^R(\Omega)$ obey well known orthogonality relations:
\begin{equation}
\int d\Omega D^R_{k,l}(\Omega)D^{R^{\prime}}_{m,n}(\Omega^{\dagger})=d^{-1}(R)\delta_{R,R^{\prime}}\delta_{k,n}\delta_{l,m}
\label{ortho}
\end{equation}

Substitute eqn (\ref{ortho}) into the l.h.s. of eq.(\ref{limitmatrix}) to obtain the Wilsonloop as in eq.(\ref{puregauge}), multiplied with the phase factor \linebreak $\exp{\{i(s_2-s_1)(C_2(R_0)-{1\over 2}TrH_0^2)/2I\}}$.

The matrix element of the propagator can be written in path integral form
\begin{equation}
\int_{U_1\Omega}^{(U_2\Omega} D\Omega(s)\exp{-i\int ds L}
\label{pathform}
\end{equation}
\noindent with $L$ as in eq.(\ref{lagrange}).
The final form of the Wilson loop in the irrep $R_0$ and with weight $H_0$ is then:
\begin{equation}
W=\int D\Omega \exp{\{-\oint ds TrH_0(\Omega(\nabla-ig\vec A).{d\vec x\over{ds}}\Omega^{\dagger})\}}
\end{equation}

The line integral on the r.h.s. can be easily transformed into a surface integral, which gives then eq.(\ref{diakonov}).

\end{document}